\pdfoutput=1
\documentclass[12pt,a4paper]{paper}
\usepackage[margin=0.75in]{geometry}
\usepackage[utf8]{inputenc}
\usepackage{caption}
\newcommand{\beginsupplement}{
        \setcounter{table}{0}
        \renewcommand{\thetable}{S\arabic{table}}
        \setcounter{figure}{0}
        \renewcommand{\thefigure}{S\arabic{figure}}
        \setcounter{section}{0}
        \renewcommand{\thesection}{S\arabic{section}}
        \renewcommand{\thesubsection}{S\arabic{section}.\arabic{subsection}}
        \setcounter{equation}{0}
        \renewcommand{\theequation}{S\arabic{equation}}
     }
\newcommand{\indep}{\raisebox{0.05em}{\rotatebox[origin=c]{90}{$\models$}}}
\usepackage{multirow}
\usepackage{amsmath}
\usepackage{enumitem} 
\usepackage{array}
\usepackage{graphicx}
\usepackage{tablefootnote}
\usepackage{lscape}
\usepackage{setspace}
\newcolumntype{P}[1]{>{\centering\arraybackslash}p{#1}}
\usepackage{authblk}

\usepackage{fancyhdr}
\begin{document}
\title{\vspace{-3cm} \huge \centering Measurement Error in Continuous Endpoints in Randomised Trials: Problems and Solutions}
\author[1, *]{L. Nab}
\author[1]{R.H.H. Groenwold}
\author[2]{P.M.J. Welsing}
\author[1]{M. van Smeden}
\affil[1]{Department of Clinical Epidemiology, Leiden University Medical Center, Leiden, Netherlands}
\affil[2]{Department of Rheumatology
\& Clinical Immunology,
University Medical Center
Utrecht, Utrecht, Netherlands}
\affil[*]{Correspondence: L.Nab@lumc.nl \authorcr
\textbf{Data availability:} The data and code used for the simulation study have been made publicly. The data is available at doi.org/10.6084/m9.figshare.7068695 and the code is available at doi.org/10.6084/m9.figshare.7068773.
\authorcr
\textbf{Funding:} This work was supported by the Netherlands Organisation for Scientific Research (NWO, project 917.16.430). }
\begin{singlespacing}
\maketitle
\end{singlespacing}
\thispagestyle{empty}

\begin{abstract}
In randomised trials, continuous endpoints are often measured with some degree of error. This study explores the impact of ignoring measurement error, and proposes methods to improve statistical inference in the presence of measurement error. Three main types of measurement error in continuous endpoints are considered: classical, systematic and differential. For each measurement error type, a corrected effect estimator is proposed. The corrected estimators and several methods for confidence interval estimation are tested in a simulation study. These methods combine information about error-prone and error-free measurements of the endpoint in individuals not included in the trial (external calibration sample). We show that if measurement error in continuous endpoints is ignored, the treatment effect estimator is unbiased when measurement error is classical, while Type-II error is increased at a given sample size. Conversely, the estimator can be substantially biased when measurement error is systematic or differential. In those cases, bias can largely be prevented and inferences improved upon using information from an external calibration sample, of which the required sample size increases as the strength of the association between the error-prone and error-free endpoint decreases. Measurement error correction using already a small (external) calibration sample is shown to improve inferences and should be considered in trials with error-prone endpoints. Implementation of the proposed correction methods is accommodated by a new software package for R.
\end{abstract}

\section{Introduction}
In randomised controlled trials, continuous endpoints are often measured with some degree of error. Examples include trial endpoints that are based on self-report (e.g. self-reported physical activity levels \cite{Cerin2016CorrelatesActivity.}), endpoints that are collected as part of routine care (e.g. in pragmatic trials \cite{Lauer2013TheResearch}), endpoints that are assessed without blinding the patient or assessor to treatment allocation (e.g. in surgical \cite{Boutron2004BlindingTrials.} or dietary \cite{Staudacher2017TheInterventions.} interventions) and an alternative endpoint assessment that substitutes a gold-standard measurement because of monetary or time constraints or ethical considerations (e.g. food frequency questionnaire as substitute for doubly-labelled water to measure energy intake \cite{Mahabir2006CalorieWomen.}). In these examples, the continuous endpoint measurements contain error in the sense that the recorded endpoints do not unequivocally reflect the endpoint one aims to measure.\\
\\
Despite calls for attention to the issue of measurement error in endpoints (e.g.  \cite{Senn2009MeasurementStatisticians}), developments and applications of correction methods for error in endpoints are still rare \cite{Keogh2016StatisticalTrials.}. Specifically, methodology that allow for correction of study estimates for the presence of measurement error have so far largely been focused on the setting of error in explanatory variables, which may give rise to inferential errors such as regression dilution bias \cite{Buonaccorsi2010MeasurementApplications,Brakenhoff2018RandomFactors.,Carroll2006,Fuller1987MeasurementModels,Gustafson2004MeasurementAdjustments.,Hutcheon2010RandomBias}. In addition, the application of correction methods for measurement errors in the applied medical literature is unusual \cite{Brakenhoff2018MeasurementReview, Shaw2018}.\\ 
\\
We provide an exploration of problems and solutions for measurement error in continuous trial endpoints. For illustration of the problems and solutions for measurement error in continuous endpoints we consider one published trial that examined the efficacy and tolerability of low-dose iron-supplements during pregnancy \cite{Makrides2003Efficacy13}. To test the effect of the iron supplementation on maternal haemoglobin levels, haemoglobin concentrations were measured at delivery in venous blood.\\
\\
This paper describes a taxonomy of measurement errors in trial endpoints, evaluates the effect of measurement errors on the analysis of trials and tests existing and proposes new methods evaluating trials containing measurement errors. Implementation of the proposed measurement error correction methods (i.e. the existing and novel methods) are supported by introducing a new R package  \texttt{mecor}, available at: www.github.com/LindaNab/mecor. This paper is structured as follows. In section \ref{sec:illex} we revisit the example trial introduced in the previous paragraph. Section \ref{sec:mestruc} presents an exploration of the influence of measurement error structures and their impact on inferences of trials. In section \ref{sec:cormeth} measurement error correction methods are proposed. A simulation study investigating the efficacy of the correction methods is presented in section \ref{sec:sim}. Conclusions and recommendations resulting from this study are provided in section \ref{sec:dis}.

\section{Illustrative example: measurement of haemoglobin levels}\label{sec:illex}
Makrides et al. \cite{Makrides2003Efficacy13} tested the efficacy of a 20-mg daily iron supplement (ferrous sulfate) on maternal iron status in pregnant women in a randomized, two-arm, double-blind, placebo-controlled trial. Respectively, 216 and 214 women were randomized to the iron supplement and placebo arm. At delivery, a 5-mL venous blood sample was collected from the women to assess haemoglobin levels as a marker for their iron status. Haemoglobin levels of women in the iron supplement arm were significantly higher than haemoglobin levels of women in the placebo arm (mean difference 6.9 CI (4.4; 9.3)). Haemoglobin concentrations were measured spectrophotometrically. Mean haemoglobin values were 137 (SD 3.2) g/L when measured by certified measurements, compared to mean 135 (SD 0.96) g/L when measured using the equipment used in the trial to measure haemoglobin levels. This might indicate small measurement errors in the measured haemoglobin levels of the women in the trial. The authors did not discuss if and how the remaining measurement error if and how could have affected their results.\\
\\
In this domain, similar trials have been conducted in which the endpoint was assessed with lower standards. For instance, in field trials testing the effectiveness of iron supplementation, capillary blood samples instead of venous blood samples are often used to measure haemoglobin levels (e.g. \cite{Zlotkin2001RandomizedAnemia.}). While easier to measure, capillary haemoglobin levels are less accurate than venous haemoglobin levels \cite{Patel2013CapillaryDonors}. We now discuss how measurement errors in haemoglobin levels might affect trial inference, by assuming hypothetical differences between capillary and venous haemoglobin levels. Two more illustrative examples are discussed in supplementary materials section 1.

\subsection{Simulations based on example trial}
We expand on the preceding example to hypothetical structures of error in measurement of the endpoints by simulation. These structures are only explained intuitively (explicit definitions are provided in section \ref{sec:mestruc}). For this example, we take the observed mean difference in haemoglobin levels in the two groups of the iron supplementation trials as a reference (6.9 g/L higher in the iron-supplemented group), and assume that haemoglobin levels are normally distributed with equal variance in both groups (SD 12.6 g/L). Fifty-thousand simulation samples were taken with 54 patients in each treatment arm. The number of patients differed from the 430 patients in the original trial to yield a Type-II error of approximately 20\% in the absence of measurement error at the usual alpha level (5\%). Treatment effect for each simulation sample (mean difference in haemoglobin levels between the two arms) was estimated by OLS regression. 

\subsubsection{Classical measurement error in example trial}
In the context of measurement of haemoglobin levels, random variability in the haemoglobin levels of capillary blood samples may be expected to vary more than haemoglobin levels in venous blood \cite{Patel2013CapillaryDonors}, independently of the true haemoglobin level and allocated treatment. Increased Type-II error is a well-known consequence of endpoints measured by the lower standard that are unbiased but more variable than the endpoints measured by the preferred measurement instruments \cite{Hutcheon2010RandomBias}. This form of measurement error is commonly described as “random measurement error” or “classical measurement error” \cite{Carroll2006}. To simulate such independent variation, we arbitrarily increased the standard deviation of haemoglobin levels by 75\% (from 12.6 to 22.05). This is equivalent to adding a term drawn from a normal distribution with mean 0 and standard deviation 18.1 to each endpoint. The impact of this imposed classical error was an increased between-replication variance of the estimated treatment effects of approximately 55\% (left plot in panel b, Figure \ref{fig:me}). The average estimated effect across simulations (depicted by the dashed line) is approximately equal to the true effect (depicted by the solid line), suggesting the classical measurement error did not introduce a bias in the estimated treatment effect (a formal proof is given in section \ref{sec:mestruc:clasme}). Type-II error increased (to 38\%) (grey area in Figure \ref{fig:me}, panel b) while Type-I error remained at the nominal level (at 5\%, illustrated by the red area in Figure \ref{fig:me}, panel b).

\subsubsection{Systematic measurement error in example trial}
It may alternatively be assumed that capillary haemoglobin levels are systematically different from venous haemoglobin levels.
This systematic difference can be either additive or multiplicative. For additive systematic measurement error, the capillary haemoglobin levels differ from venous haemoglobin levels with a certain constant, independently of venous haemoglobin levels. This implies that in both treatment groups mean haemoglobin level is higher, but that the difference between the two treatment groups is unbiased. The term systematic measurement error is often used to indicate multiplicative measurement error \cite{Keogh2014}. In that case, the expected capillary haemoglobin levels are equal to venous haemoglobin levels multiplied by a certain constant. Consequently, haemoglobin levels in capillary blood are more accurately measured in patients with low venous haemoglobin levels than in patients with high true haemoglobin levels (or vice versa). Under the assumption of a non-zero treatment effect, the expected difference between mean haemoglobin levels between the two treatment groups is biased; in the absence of a treatment effect, the expected difference between the two groups will remain unaffected. To simulate, we assumed that capillary haemoglobin levels are 1.05 times haemoglobin levels and we increased the standard deviation of haemoglobin levels by 75\%, equivalent to the previous example. The impact of this imposed systematic measurement error structure is that the average treatment effect was biased, increasing from 6.9 to 7.2, and that there is an increased between-replication variance of the estimated treatment effect of approximately 66\% (left plot in Figure \ref{fig:me}, panel c). Type-II error increased (to 37\%) (grey area in Figure \ref{fig:me}, panel c) while Type-I error remained at rate close to nominal level (at 5\%) (red area in Figure \ref{fig:me}, panel c).

\subsubsection{Differential measurement error in example trial}
\begin{figure}
\centering
\includegraphics{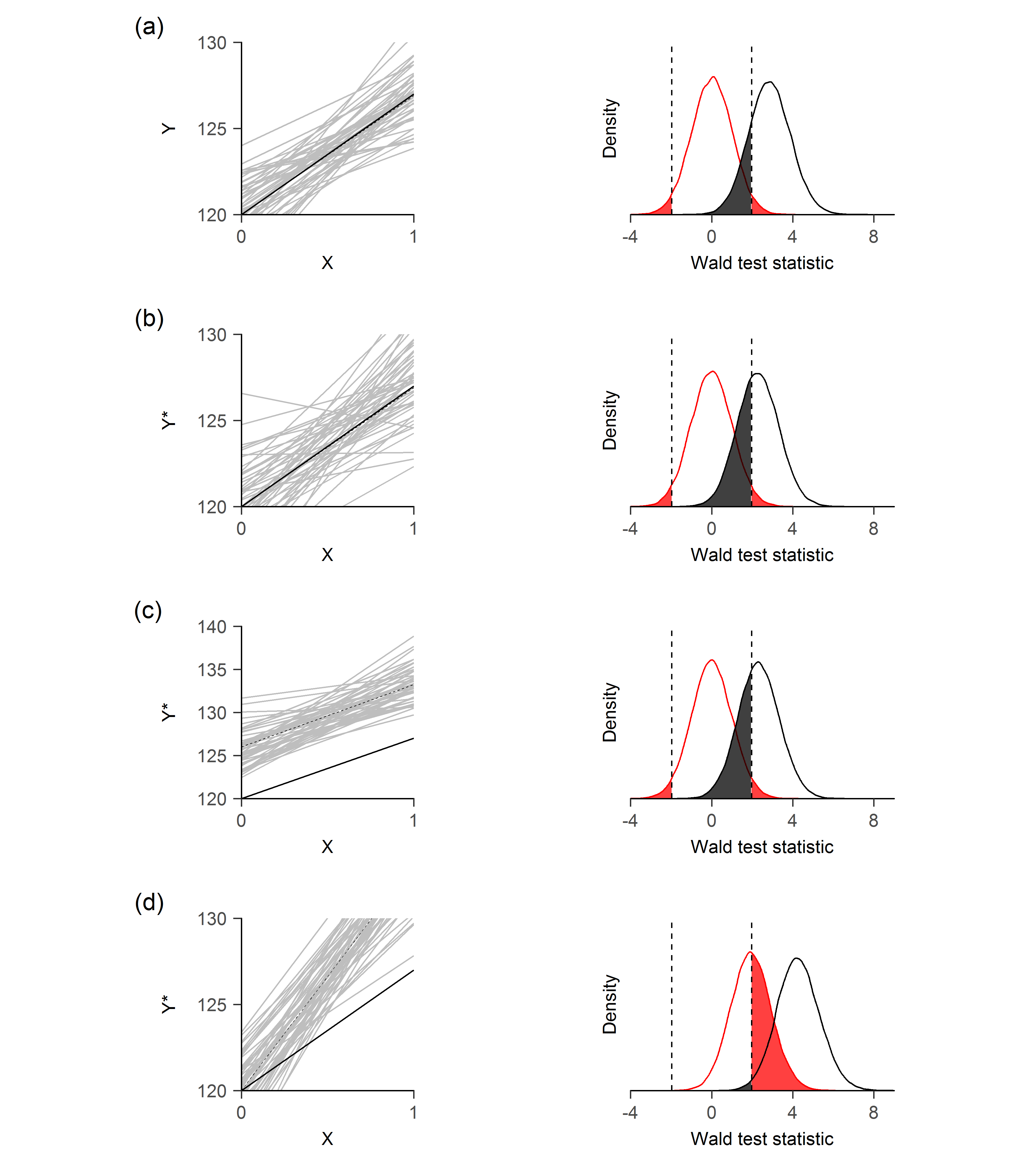}
\caption{Illustration of impact of hypothetical measurement error in example trial 1 \cite{Makrides2003Efficacy13}: (a) no measurement error; (b) classical measurement error; (c) systematic measurement error; (d) differential measurement error. The left plots depict every thousandth estimated OLS regression line (grey lines), the average estimated treatment effect (dashed line) and the true effect (black line). The right plots depict the density distribution of the Wald test-statistic of the slope of the regression line, under the null hypothesis of no effect (red distribution) and the alternative hypothesis of any effect (black distribution).}\label{fig:me}
\end{figure}

The measurement error (structure) may also differ between the treatment arms. In an extreme scenario, haemoglobin levels in placebo group patients would be measured by venous blood samples while patients in active arm (iron supplemented) would be measured using capillary blood samples. To simulate such a scenario, we assume the same systematic error structure from the previous paragraph, now only applying to the active group. Additionally, we assume classical measurement error in the placebo group. This scenario classifies as differential measurement error \cite{Keogh2016StatisticalTrials.}. The impact of this measurement error structure is that the average treatment effect was biased, increasing from 6.9 to 13.3, and that the between-replication variance of the estimated treatment effect is increased by approximately 62\% (left plot in Figure \ref{fig:me}, panel d). Type-II error decreased (to 0.1\%) (grey area in Figure \ref{fig:me}, panel d) and Type-I error rates increased (to 48\%) (red area in Figure \ref{fig:me}, panel d).\\

\section{Measurement error structures}\label{sec:mestruc}
Consider a two-arm randomized controlled trial that compares the effects of two treatments ($X \in \{0,1\}$), where 0 may represent a placebo treatment or an active comparator. Let $Y$ denote the true (or preferred) trial endpoint and $Y^*$ an error prone operationalisation of $Y$. We will assume that both $Y$ and $Y^*$ are measured on a continuous scale. We assume a linear regression model for the endpoint $Y$:
\begin{equation}\label{eq:gen}
Y=\alpha_Y + \beta_YX + \varepsilon,
\end{equation}
where $\varepsilon$ is iid normally distributed with mean 0 and variance $\sigma^2$. 
Under these assumptions and assumptions about the model for $Y^*$ (described below), simple formulas for the bias in the OLS estimator of the treatment effect can be derived. Details of these derivations can be found in the supplementary materials, section 2.
\subsection{Classical measurement error}\label{sec:mestruc:clasme}
There is classical measurement error in $Y^*$ if $Y^*$ is an unbiased proxy for $Y$ \cite{Carroll2006}: $Y^*=Y+e$,
where $e$ has mean 0 and Var$(e)=\tau^2$ and $e$ independent of $Y$, $X$, $\varepsilon$ in (\ref{eq:gen}). Using $Y^*$ instead of $Y$ in the linear model yields:
\begin{equation}\label{eq:lm}
Y^* = \alpha_Y^* + \beta_Y^*X + \delta,
\end{equation}
Where $\beta_Y^*=\beta_Y$ and the residuals $\delta$ have mean 0 and variance $\sigma_{\delta}^2 = \sigma^2 + \tau^2$. This leads to a larger variance in $\hat{\beta}_Y^*$ (the estimator for $\beta_Y^*$) compared to the variance in $\hat{\beta}_Y$ (the estimator for $\beta_Y$). Consequently, classical measurement error will not lead to bias in the effect estimator but will increase Type-II for a given sample size.
\subsection{Heteroscedastic measurement error}\label{sec:mestruc:hetme}
In the above we assumed that the variance in $e$ is equal in both arms. When this assumption is violated, there is so called heteroscedastic measurement error. Heteroscedastic error will not lead to bias in the effect estimator, but will invalidate the estimator of the variance of $\hat{\beta}_Y^*$ (proof is given in supplementary materials section 2).

\subsection{Systematic measurement error}\label{sec:sysme}
There is systematic measurement error in $Y^*$ if $Y^*$ depends systematically on $Y$: $Y^* = \theta_0 + \theta_1 Y + e$, where $e$ has mean 0 and Var$(e)=\tau^2$ and $e$ independent of $Y$, $X$, $\varepsilon$ in (\ref{eq:gen}). Throughout, we assume systematic measurement error if $\theta_0 \neq 0$  or $\theta_1 \neq 1$ (and of course, $\theta_1 \neq 0$  in all cases). We assume independence between $e$ and $Y$, $X$, $\varepsilon$ in (\ref{eq:gen}). Using $Y^*$ with systematic measurement error in the linear model yields in the model defined by (\ref{eq:lm}) where $\beta_Y^*=\theta_1\beta_Y$ and the residuals $\delta$ have mean 0 and variance $\sigma_\delta=\theta_1^2\sigma^2 + \tau^2$. Depending on the value of $\theta_1$, the variance of $\hat{\beta_Y^*}$ is larger or smaller than the variance of $\hat{\beta_Y}$. Hence, Type-II error will either decrease or increase under systematic measurement. Type-I error is unaffected since if $\beta_Y = 0$, $\beta_Y^* = 0$ (i.e., tests for null effects are still valid under systematic measurement error) (proof is given in supplementary materials section 2).

\subsection{Differential measurement error}\label{sec:difme}
There is differential measurement error in $Y^*$ if $Y^*$ depends systematically on $Y$ varying for $X$: $Y^*=\theta_{00}+(\theta_{01}-\theta_{00})X+\theta_{10}Y+(\theta_{11}-\theta_{10})XY+ e_X$, where $e_X$ has mean 0 and Var$(e)=\tau_X^2$ and $e_X$ independent of $Y$, and $\varepsilon$ in (\ref{eq:gen}) for $X=0,1$. Using $Y^*$ with differential measurement error in the linear model yields in the model defined in (\ref{eq:lm}) where $\beta_Y^*= \theta_{01}-\theta_{00}+(\theta_{11}-\theta_{10})\alpha_Y+\theta_{11}\beta_Y$ and the residuals $\delta$ have mean 0 and variance $\big[\theta_{10}^2 + (\theta_{11}^2-\theta_{10}^2)X\big]\sigma^2+\tau_X^2$ for $X = 0, 1$. Since the residual variance is not equal in both arms, the estimator of the variance of $\hat{\beta}_Y^*$ is invalid, and will underestimate the true variance. A heteroscedastic consistent estimator of the variance of $\hat{\beta}_Y^*$ is provided by the White estimator \cite{Long2000UsingModel}. Assuming that the White estimator is used to estimate the variance of $\hat{\beta}_Y^*$, Type-I error is not expected the nominal level ($\alpha$) and Type-II error will decrease or increase under the differential measurement error model (proof is given in supplementary materials section 2). 

\section{Correction methods for measurement error in a continuous trial endpoint}\label{sec:cormeth}
In this section we describe several approaches to address measurement error in the trial endpoint. Throughout, we assume that $Y^*$ is measured for all $i = 1,\hdots,N$ randomly allocated patients in the trial. We also assume that $Y$ and $Y^*$ are both measured for a smaller set of different individuals not included in the trial ($j = 1,\hdots,K, K < N$), hereinafter referred to as the external calibration sample. In all but one case, it is assumed that only $Y^*$ and $Y$ are measured in the external calibration sample. In the case that the error in $Y^*$ is different for the two treatment groups, it is assumed that the external calibration sample is in the form of a small pilot study where both treatments are allocated (i.e., $Y^*$ and $Y$ are both measured after assignment of $X$). Instead of external calibration data, we could use internal calibration data to correct for measurement error ($Y$ and $Y^*$ are both measured in a small subset of the trial), which is not considered in this paper as it was studied elsewhere \cite{Keogh2016StatisticalTrials.}.\\
\\
A well-known consequence of classical measurement error in a continuous trial endpoint is that a larger sample size (as compared to the same situations without the measurement error) is needed to compensate for the reduced precision \cite{Hutcheon2010RandomBias}. For example, the new sample size $N^*$ may be calculated by $N/R$ formula where $R$ is the reliability coefficient and $N$ the original sample size for the trial \cite{Fitzmaurice2002MeasurementReliability}. For solutions for heteroscedastic measurement error, we refer to standard theory of dealing with heteroscedastic errors in regression to find an unbiased estimator for the variance of $\hat{\beta}_{Y^*}$ (e.g. see \cite{Long2000UsingModel} for an overview of different heteroscedasticity consistent covariance matrices).\\ 
\\
Hereinafter we focus on measurement error in $Y^*$ that is either systematic or differential, both of which have been shown to introduce bias in the effect estimator if measurement error is neglected (section \ref{sec:mestruc}). Consistent estimators for the intervention effects are introduced, and various methods for constructing confidence intervals for these estimators are discussed. Section 3 in the supplementary materials provides an explanation of the results stated in this section. Throughout, we assume that $Y^*$ is measured for all $i = 1,\hdots,N$ patients in the trial. We also assume that $Y$ and $Y^*$ are both measured for a smaller set of different individuals not included in the trial ($j = 1,\hdots,K, K < N$), hereinafter referred to as the external calibration sample. For an earlier exploration of the use of an internal calibration set when there is systematic or differential measurement error in endpoints, see \cite{Keogh2016StatisticalTrials.}. 

\subsection{Systematic measurement error}\label{sec:syscorest}
From section \ref{sec:sysme} it follows that natural estimators for $\alpha_Y$ and $\beta_Y$ are
\begin{align}\label{eq:estsme}
\hat{\alpha}_Y=(\hat{\alpha}_{Y^*}-\hat{\theta}_0)/\hat{\theta}_1  \text{\quad and \quad}
\hat{\beta}_Y=\hat{\beta}_{Y^*}/\hat{\theta}_1,
\end{align}
Where $\hat{\theta}_0$ and $\hat{\theta}_1$ are the estimated error parameters from the calibration data set using standard OLS regression. From equation (\ref{eq:estsme}), it becomes apparent that $\hat{\theta}_1$ needs to be assumed bounded away from zero for finite estimates of  $\hat{\alpha}_Y$  and $\hat{\beta}_Y $ \cite{Buonaccorsi2010MeasurementApplications}. The estimators in (\ref{eq:estsme}) are consistent, see for a proof section 3.1 in the supplementary materials.\\
The variance of the estimators defined in (\ref{eq:estsme}) can be approximated using the Delta method \cite{Buonaccorsi1991MeasurementMeans}, the Fieller method \cite{Buonaccorsi1991MeasurementMeans}, the Zero-variance method and by bootstrap \cite{Efron1979BootstrapJackknife}. Further details are provided in section 3.1 of the supplementary materials.

\subsection{Differential measurement error}\label{sec:difcorest}
From section \ref{sec:difme} it follows that natural estimators for $\alpha_Y$ and $\beta_Y$ are, 
\begin{equation}\label{eq:estdme}
\hat{\alpha}_Y=(\hat{\alpha}_{Y^*}-\hat{\theta}_{00})/\hat{\theta}_{10} \textrm{\quad and \quad}
\hat{\beta}_Y=(\hat{\beta}_{Y^*}+\hat{\alpha}_{Y^*}-\hat{\theta}_{01})/\hat{\theta}_{11}-\hat{\alpha}_{Y},
\end{equation}
where $\hat{\theta}_{00}$, $\hat{\theta}_{10}$, $\hat{\theta}_{01}$ and $\hat{\theta}_{11}$ are estimated from the external calibration set using standard OLS estimators. Here it is assumed that both $\hat{\theta}_{10}$ and $\hat{\theta}_{11}$ are bounded away from zero (for reasons similar to those mentioned in section \ref{sec:syscorest}). The estimators in (\ref{eq:estdme}) are consistent, see for a proof section 3.1 of the supplementary materials.
The variance of the estimators defined in (\ref{eq:estdme}) can be approximated using the Delta method \cite{Buonaccorsi1991MeasurementMeans}, the Zero-variance method and by bootstrap \cite{Efron1979BootstrapJackknife}. Further details are provided in section 3.2 of the supplementary materials.

\section{Simulation study}\label{sec:sim}
The finite sample performance of the measurement error corrected estimators of the treatment effect was studied by simulation. We focussed on the situation of a two-arm trial in which the continuous surrogate endpoint $Y^*$ was measured with systematic or differential measurement error, and in which an external calibration set was available, which was varied in size. The results from example trial 1 are used to motivate our simulation study (see section \ref{sec:illex}).

\subsection{Data generation}
Data were generated for a sample of $N = 400$ individuals, approximately equal to the size of example trial 1 \cite{Makrides2003Efficacy13}. The individuals were equally divided in the two treatment arms. The true endpoints were generated according to model (\ref{eq:gen}), assuming iid normal errors, and using the estimated characteristics found in example trial 1 ($\alpha_Y=120$, $\beta_Y=6.9$ and $\sigma=12.6$). Surrogate endpoints $Y^*$ were generated under models for systematic measurement error and differential measurement error described in section \ref{sec:sysme} and \ref{sec:difme}, respectively. \\
\\
For systematic measurement error in $Y^*$, we set $\theta_0=0$ and $\theta_1=1.05$. Under the differential measurement error model we set $\theta_{00}=0$, $\theta_{01}=0$, $\theta_{10}=1$, $\theta_{11}=1.05$. We considered three scenarios based on the coefficient of determination between the $Y^*$ and $Y$, $R^2_{Y^*,Y}$: (i) $R^2_{Y^*,Y}=0.8$, (ii) $R^2_{Y^*,Y}=0.5$ and (iii) $R^2_{Y^*,Y}=0.2$. This large range in coefficient of determination values reflects the wide variation we anticipate in practice from very strong correlations between $Y*$ and $Y$ ($R^2_{Y^*,Y}$ = 0.8) to weak correlations ($R^2_{Y^*,Y}$ = 0.2), as for example, one could expect in the context of trials with dietary intake as endpoints \cite{Keogh2016StatisticalTrials.,Freedman2018}. For $R^2_{Y^*,Y}=0.8$, $\tau=6.6$ for systematic measurement error and $\tau_0 = 6.3$ and $\tau_1=6.6$  for differential measurement error. For $R^2_{Y^*,Y}=0.5$, $\tau=13.2$ for systematic measurement error and $\tau_0 = 12.6$ and $\tau_1=13.2$  for differential measurement error. For $R^2_{Y^*,Y}=0.2$, $\tau=26.5$ for systematic measurement error and $\tau_0 = 25.2$ and $\tau_1=26.5$  for differential measurement error. Additionally, we considered a scenario with greater systematic measurement error holding $\theta_0=0$ and $\theta_1=1.25$. Here, we only studied a high coefficient of determination $R^2_{Y^*,Y}=0.8$, implying that $\tau=7.9 $. \\
\\
For the scenarios with systematic measurement error induced, a separate calibration set was generated of size $K$ with the characteristics of the placebo arm for each simulated data set. For differential measurement error scenarios, a calibration data set was generated of size $K$ for each simulated data set, with $K_0=K_1=K/2$ subjects equally divided over the two treatment groups. The sample size of the external calibration data set $(K)$ was varied with $K \in \{5,7,10,15,20,30,40,50\}$ for systematic measurement error and $K \in \{10,20,30,40,50\}$ for differential measurement error. 

\subsection{Computation}\label{sec:sim:comp}
For each simulated data set the corrected treatment effect estimator (\ref{eq:estsme}) for systematic error and (\ref{eq:estdme}) for differential error were applied. In systematic measurement error scenarios, confidence intervals for the corrected estimator for $\alpha=0.05$ were constructed by using the Zero-variance method, the Delta method, the Fieller method, and the Bootstrap method based on 999 replicates (as defined in section \ref{sec:syscorest}). In the case of differential measurement error, confidence intervals for the corrected estimator for $\alpha=0.05$ were constructed by using the Zero-Variance method, the Delta method and the Bootstrap method based on 999 replicates (as defined in section 4.2). The HC3 heteroscedastic consistent variance estimator was used to accommodate for heteroscedastic error in the differential measurement error scenario \cite{Long2000UsingModel}. Furthermore, for both the systematic and differential measurement error scenarios the naive analysis was performed (resulting in a naive effect estimate and naive confidence interval), which is the 'regular' analysis which would be performed if measurement errors were neglected.\\
\\
We studied performance of the corrected treatment effect estimators in terms of percentage bias \cite{Burton2006TheStatistics}, empirical standard error (EmpSE) and square root of the  mean squared error (SqrtMSE) \cite{Morris2019}. The performance of the methods for constructing the confidence intervals was studied in terms of coverage and Type-II error \cite{Morris2019}.\\
\\
In our simulations, the Fieller method resulted in undefined confidence intervals if in an iteration $\hat{\theta}_1/\sqrt{t^2/S^{(c)}_{yy}}>t_{N-2}$. The percentage of iterations for which the Fieller method failed to construct confidence intervals is reported. If the Fieller method resulted in undefined confidence intervals in more than 5\% of cases in one simulation scenario, the coverage and average confidence interval width were not calculated as this would result in unfair comparisons between the different confidence interval constructing methods. The bootstrap confidence intervals were based on less than 999 estimates in case the sample drawn from the external calibration set consisted of $K$ equal replicates. These errors occurred more frequently for small values of $K$ and low R-squared. All simulations were run in R version 3.4, using the library \texttt{mecor} (version 0.1.0). The results of the simulation are available at doi.org/10.6084/m9.figshare.7068695 and the code is available at doi.org/10.6084/m9.figshare.7068773, together with the seed used for the simulation study.   

\subsection{Results of simulation study}
\subsubsection{Systematic measurement error}
Table \ref{tbl:sme} shows percentage bias, EmpSE and SqrtMSE of the naive estimator and the corrected estimator for $\theta_1=1.25$ and $R^2_{Y^*,Y}=0.8$ and $\theta_1=1.05$ and $R^2_{Y^*,Y}=0.8$, $R^2_{Y^*,Y}=0.5$ and $R^2_{Y^*,Y}=0.2$ and $K \in \{5,7,10,15,20,30,40,50\}$ when there is systematic measurement error. Naturally, the percentage of bias in the naive estimator is about 5\% if $\theta_1 = 1.05$ and 25\% if $\theta_1=1.25$. For the corrected estimator and $\theta_1 = 1.05$ or $\theta_1=1.25$ and $R^2_{Y^*,Y}=0.8$, percentage bias, EmpSE and SqrtMSE of $\hat{\beta}_Y$ are reasonably small for $K \geq 10$. Yet, as the bias in the naive estimator is small when $\theta_1=1.05$, SqrtMSE of the corrected estimator is never lower than the SqrtMSE of the naive estimator. However, if bias in the naive estimator is greater ($\theta_1=1.25$), SqrtMSE of the corrected estimator is smaller than SqrtMSE of the naive estimator for $K \geq 15$. For the corrected estimator and $\theta_1 = 1.05$ and $R^2_{Y^*,Y}=0.5$, bias is reasonably small for $K \geq 30$. Nevertheless, SqrtMSE of the corrected estimator is always greater than SqrtMSE of the naive estimator. For the corrected estimator and $\theta_1 = 1.05$ and $R^2_{Y^*,Y}=0.2$, bias of $\hat{\beta}_Y$ fluctuates and EmpSE and SqrtMSE is large for all $K's$. The estimates of the intervention effect using the corrected estimator of each 10th iteration of our simulation is shown in Figure 2, which provides a clear visualization of the results formerly discussed. The bigger the sample size of the external calibration set and the higher R-squared, the better the performance of the corrected estimator. The sampling distribution of $\hat{\theta}_1$ depicted in Figure 3 explains why there is so much variation in the corrected effect estimator for small sample sizes of the external calibration set and low R-squared. Namely, for a number of iterations in our simulation, $\hat{\theta}_1$ was estimated close to zero, expanding the corrected estimator the same number of times resulting in large bias, EmpSE and MSE. Note that if $\hat{\theta}_1<0$, the sign of the corrected estimator changes, explaining why the corrected estimate of the intervention effect is sometimes below zero.\\
\\
For $\mathrm{R}^2_{Y^*,Y}=0.8$ and both $\theta_1 = 1.05$ and $\theta_1 = 1.25$, the Fieller method failed to construct confidence intervals in 15, 5, 1 and 0.1 \% of simulated datasets for respectively $K = 5, 7, 10, 15$. Therefore, coverage and average confidence interval width of the Fieller method is not evaluated for $K \in \{5,7\}$. For $\mathrm{R}^2_{Y^*,Y}=0.5$, the Fieller method failed to construct confidence intervals in 48, 36, 22, 8, 3, 0.3 \% of simulated data sets for $K \in \{5,7,10,15,20,30\}$, respectively. Consequently, coverage and average confidence interval width is not evaluated for $K \in \{5,7, 10, 15\}$. For $\mathrm{R}^2_{Y^*,Y}=0.2$, the Fieller method failed to construct confidence intervals in 74, 71, 64, 53, 43, 26, 15 and 8 \% of simulated data sets for $K \in \{5,7,10,15,20,30,40,50\}$, respectively (i.e., in every case more than 5\%, thus the Fieller method is not evaluated for $\mathrm{R}^2_{Y^*,Y}=0.2$).\\
\\
Table \ref{tbl:sme} shows coverage of the true intervention effect in the constructed confidence intervals and average confidence interval width using the Zero-variance, Delta, Fieller and Bootstrap method. Using Wald confidence intervals for the naive effect estimator nearly yielded 95\% coverage of the true treatment effect of 6.9, because for $\theta_1=1.05$ the bias percentage in the naive estimator is small (i.e., $5\%$). Yet, as bias percentage increased in the naive estimator for $\theta_1 = 1.25$ (i.e., $25\%$) coverage dropped to 83.5\%. The Zero-variance method yielded too narrow confidence intervals for all scenario's, an intuitively clear result as the Zero-variance method neglects the variance in $\hat{\theta}_1$. For $\mathrm{R}^2_{Y^*,Y}=0.8$ the Delta, Fieller and Bootstrap method constructed correct confidence intervals for $K\geq 15$. For $K \leq 10$ the Delta method and the Fieller method constructed too narrow confidence intervals, and the Bootstrap method too broad confidence intervals. For $\mathrm{R}^2_{Y^*,Y}=0.5$ the Delta and Bootstrap method constructed correct confidence intervals for $K\geq 30$. For $K \leq 
20$ the Delta method constructed too narrow confidence intervals, and the Bootstrap method too broad confidence intervals. Coverage of the Fieller method was about the desired 95\% level for $K \geq 30$.\\
\\
Using the naive effect estimator, Type-II error was 0.2\%, 2.9\% and 31.6\% for $\mathrm{R}^2_{Y^*,Y}=0.8$ (both for $\theta_1 = 1.05$ and $\theta_1=1.25$), $\mathrm{R}^2_{Y^*,Y}=0.5$ and $\mathrm{R}^2_{Y^*,Y}=0.2$, respectively. Type-II error in the corrected estimator using the Zero-variance, Delta and Boostrap method was 0\%. For the considered scenario's using the Fieller method, Type-II error was 0.02\% for $\mathrm{R}^2_{Y^*,Y}=0.8$ and 2.9\% for $\mathrm{R}^2_{Y^*,Y}=0.5$.

\subsubsection{Differential measurement error}
Table \ref{tbl:dme} shows percentage bias, EmpSE and SqrtMSE of the naive estimator and the corrected estimator for $R^2_{Y^*,Y}=0.8$, $R^2_{Y^*,Y}=0.5$ and $R^2_{Y^*,Y}=0.2$ and $K \in \{5,7,10,15,20,30,40,50\}$ when there is differential measurement error. The percentage bias in the naive estimator was about 92\%. For the corrected estimator and $R^2_{Y^*,Y}=0.8$, percentage bias, EmpSE and SqrtMSE of $\hat{\beta}_Y$ are reasonably small for $K \geq 20$. For the naive estimator and $R^2_{Y^*,Y}=0.5$, percentage bias, EmpSE and MSE of the corrected estimator are small for $K = 50$. For the naive estimator and $R^2_{Y^*,Y}=0.2$, percentage bias, EmpSE and MSE of the corrected estimator is large for all $K$'s. The estimates of the intervention effect using the corrected estimator of each 10th iteration of our simulation is shown in Figure 4, which provides a clear visualization of the results formerly discussed. The bigger the sample size of the external calibration set and the higher R-squared, the better the performance of the corrected estimator.\\
\\
Table \ref{tbl:dme} shows coverage of the true intervention effect in the constructed confidence intervals and average confidence interval width using the Zero-Variance, Delta and Bootstrap method. Coverage of the true treatment effect of 6.9 using Wald confidence intervals for the naive effect estimator were about 1\%, 7\% and 41\% for $R^2_{Y^*,Y}=0.8$, $R^2_{Y^*,Y}=0.5$ and $R^2_{Y^*,Y}=0.2$, respectively. In all cases, the Zero-Variance method yielded too narrow confidence intervals; the Delta method yielded too broad confidence intervals and the Bootstrap method yielded mostly too broad confidence intervals, except for $R^2_{Y^*,Y}=0.8$ and $K=30$ and $K=40$ (too narrow). For $R^2_{Y^*,Y}=0.8$ and $K=50$, coverage of the true intervention effect was 95\%.\\
\\
Type-II error in the naive effect estimator was 0\%, 0\% and 0.4\% for $\mathrm{R}^2_{Y^*,Y}=0.8$, $\mathrm{R}^2_{Y^*,Y}=0.5$ and $\mathrm{R}^2_{Y^*,Y}=0.2$, respectively. Type-II error in the corrected effect estimator using the Zero-variance, Delta and Bootstrap method was 0\%.

\subsection{Measurement error dependent on a prognostic factor}
Above, we focused on measurement errors in endpoints that are either systematic (linearly dependent on true endpoint) or differential (linearly dependent on true endpoint and exposure). Yet, measurement error could depend on prognostic factors. For example, measurement error in haemoglobin levels measured in capillary blood may differ for women and men \cite{Patel2013CapillaryDonors}. Moreover, haemoglobin levels are, on average, higher in men than women.  To illustrate the effect of measurement error that is dependent on a prognostic factor, we use example trial 1, here assuming that it was conducted in women and men. Data were generated for a sample of $N=400$ individuals, equally divided in two treatment arms and with equal sex distribution in both arms. Let the proportion of women in the sample be 75\% ($S=1$ for men and $S=0$ for women). Further, assume $Y=120+6.9X +10S+\varepsilon$, where $\varepsilon$ has mean 0 and Var$(\varepsilon)=158.8$. Additionally, assume additive systematic measurement error in $Y^*$, $Y^*=Y+0.5S+e$ (additive systematic measurement error in men and random measurement error in women), where $e$ has mean 0 and Var$(e)=6.6$ and $e$ independent of $Y$, $X$, $S$ and $\varepsilon$. In a simulation of 10,000 replicates we estimated the effect of $Y^*$ on $X$ (naive analysis) and the effect of $Y^*$ on $X$, conditional for $S$ (conditional analysis). In section 4 of the supplementary materials, we proof that both analyses will result in correct estimation of the treatment effect. The results of the simulation study show that the average treatment effect estimate of both analyses was 6.89, indicating that there is no bias in either of the analyses. Yet, the empirical variance of the effect estimate in the 10,000 replicates was somewhat lower for the conditional analysis compared to the naive analysis (2.01 vs. 2.22), indicating an efficiency gain in favor of the conditional analysis. By assuming that randomisation was well-performed, measurement error dependent on a prognostic factor does not introduce bias in the naive analysis other than the biases already discussed.


\begin{figure}
\centering
\includegraphics[scale = 0.8]{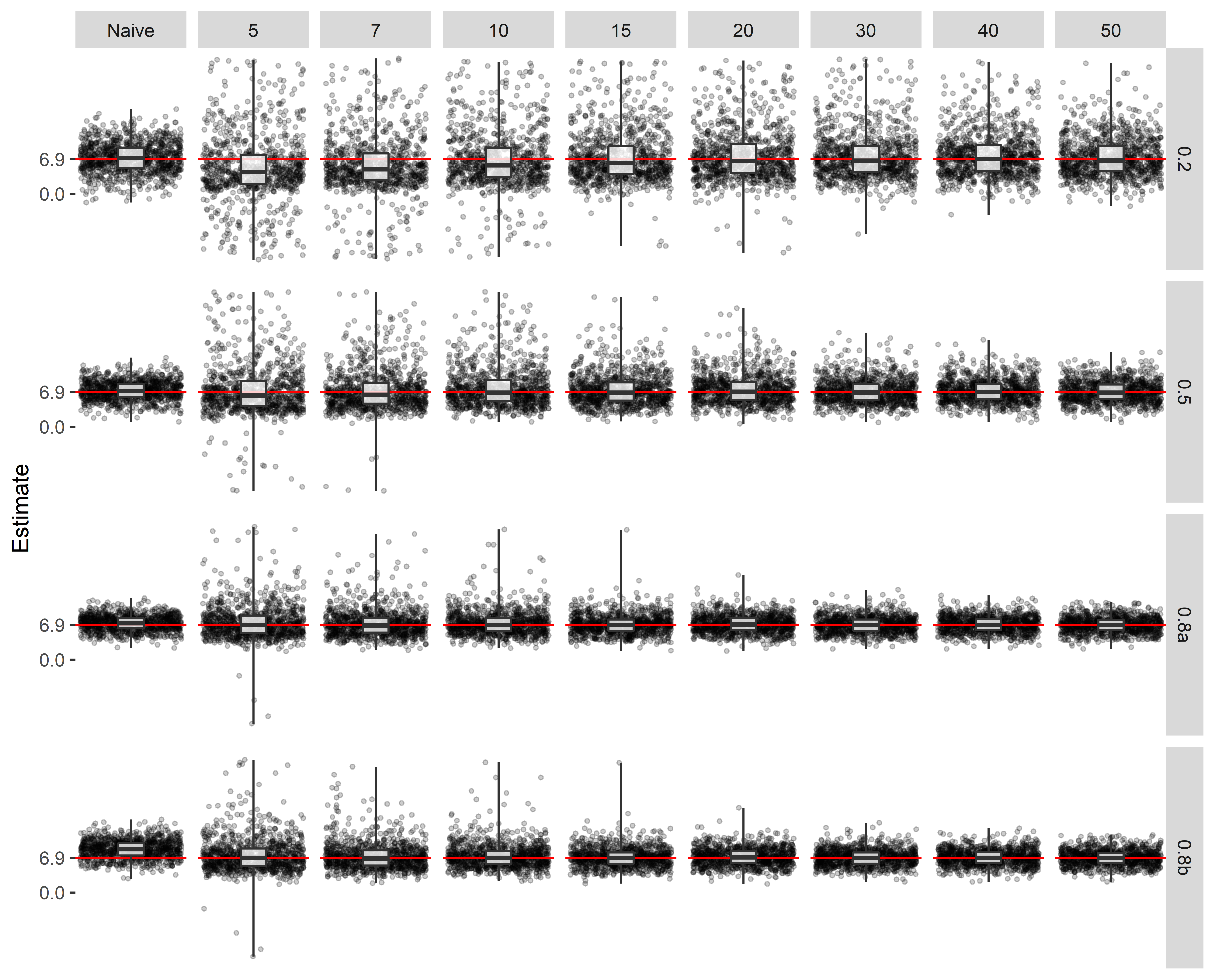}
\caption{Estimates of the treatment effect using the naive estimator and corrected estimator for different values of R-squared (row grids) and different sample sizes of the external calibration set (column grids) under systematic measurement error ($\theta_1 = 1.05$ (0.2; 0.5; 0.8a) or $\theta_1 = 1.25$ (0.8b)). Each grid is based on every 10th estimate of a simulation of 10,000 replicates, using an estimand of 6.9 (indicated by the red line), based on example trial 1 by Makrides et al. \cite{Makrides2003Efficacy13}.}\label{fig:estimatessme}
\end{figure}

\begin{figure}
\centering
\includegraphics[scale = 0.8]{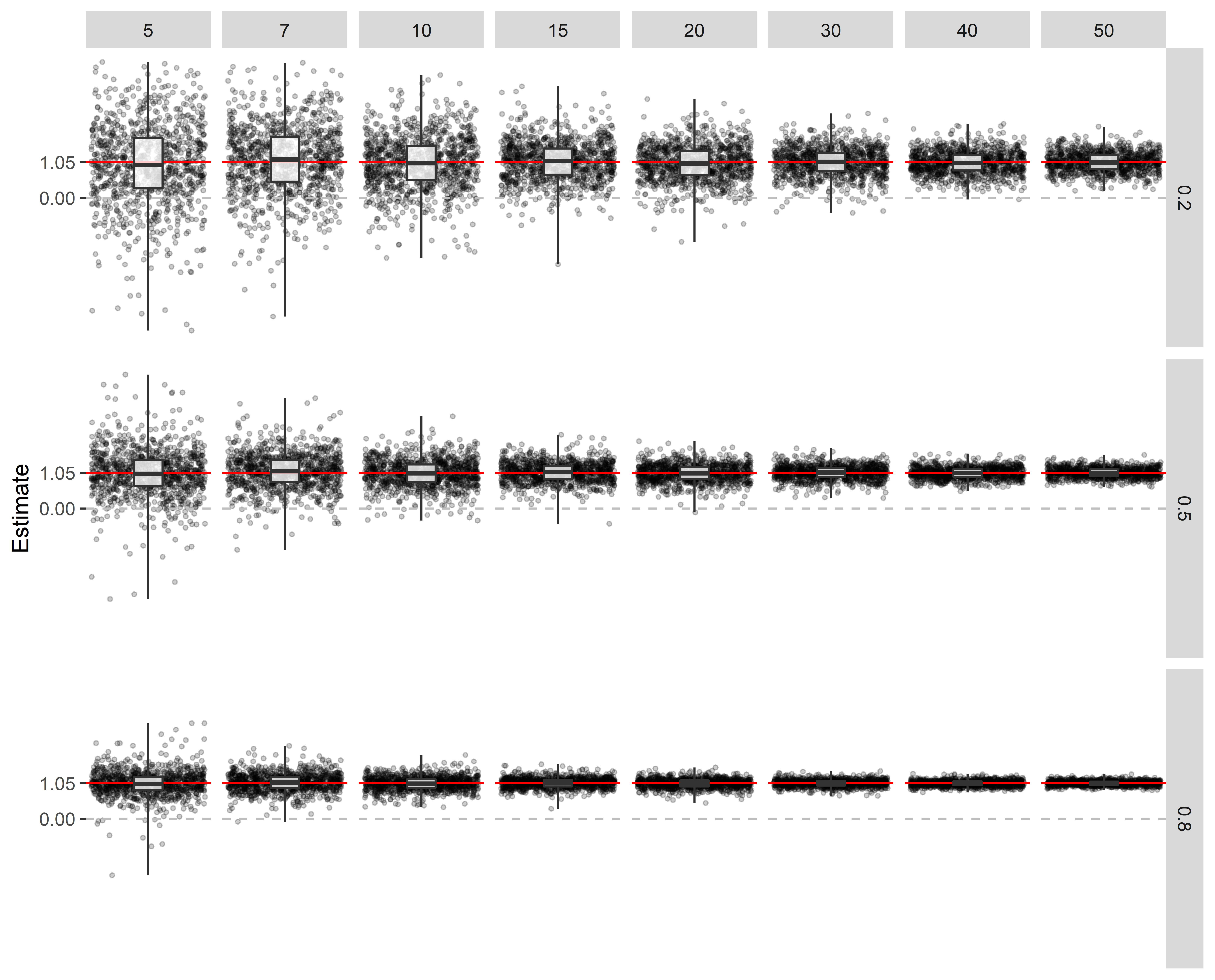}
\caption{Estimates of $\theta_1$ (i.e. slope of the systematic measurement error model) for different values of R-squared (row grids) and different sample sizes of the external calibration set (column grids). Each grid is based on every 10th estimate of a simulation of 10,000 replicates, using an estimand of 1.05 (indicated by the red line).}\label{fig:thetassme}
\end{figure}

\begin{figure}
\centering
\includegraphics[scale = 0.8]{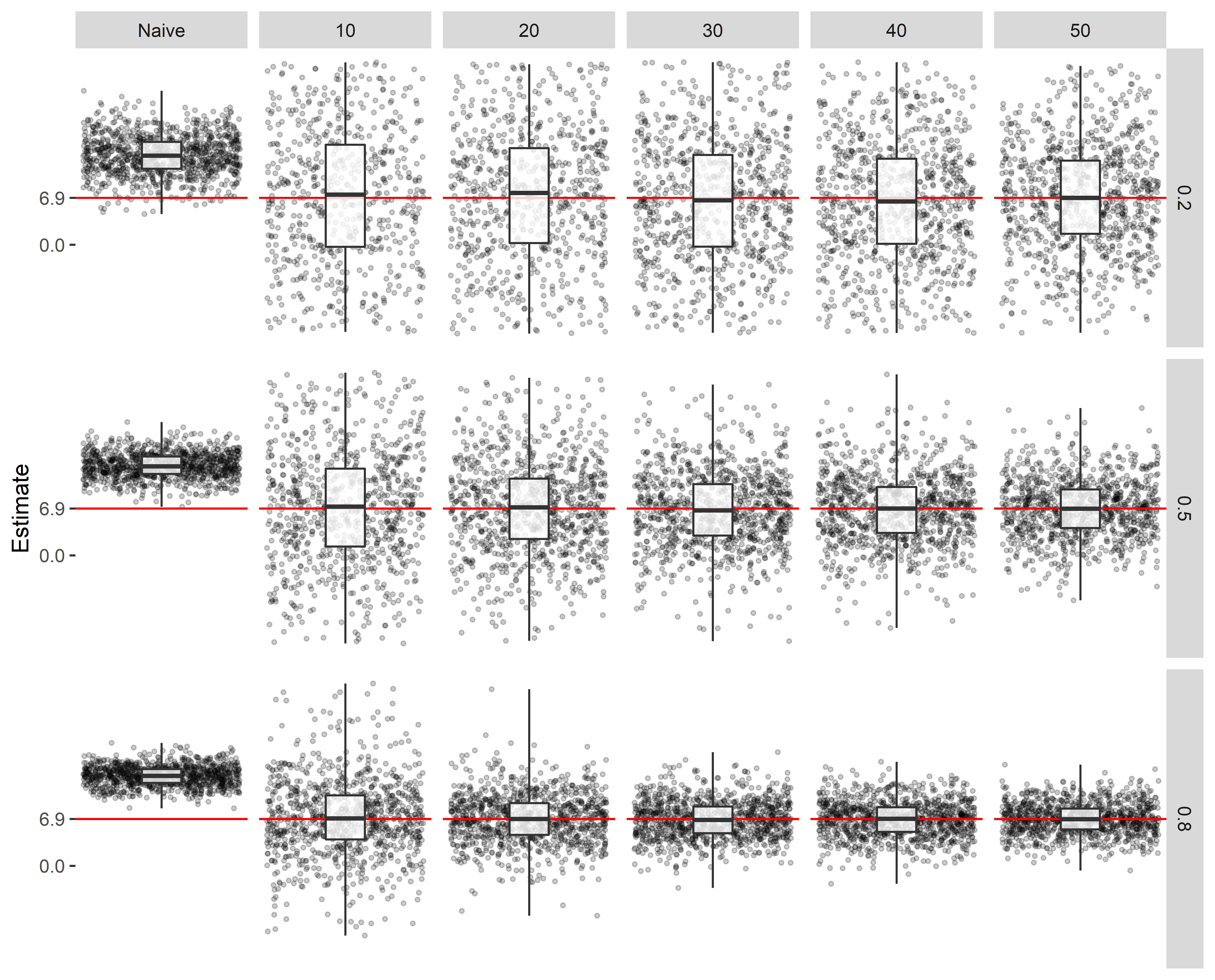}
\caption{Estimates of the treatment effect using the naive estimator and corrected estimator for different values of R-squared (row grids) and different sample sizes of the external calibration set (column grids) under differential measurement error ($\theta_{00} = 0$, $\theta_{10} = 1$, $\theta_{01} = 0$, $\theta_{11} = 1.05$). Each grid is based on every 10th estimate of a simulation of 10,000 replicates, using an estimand of 6.9 (indicated by the red line), based on example trial 1 by Makrides et al. \cite{Makrides2003Efficacy13}.}\label{fig:estimatesdme}
\end{figure}

\section{Discussion}\label{sec:dis}
This paper outlined the ramifications for randomised trial inferences when a continuous endpoint is measured with error. Our study showed that when this measurement error is ignored, not only can trial results be hampered by a loss in precision of the treatment effect estimate (i.e. increased Type-II error for a given sample size), but trial inferences can be impacted through bias in the treatment effect estimator and a null-hypothesis significance test for the treatment effect can deviate substantially from the nominal level. In this article we proposed a number of regression calibration-like correction methods to reduce the bias in the treatment effect estimator and obtain confidence intervals with nominal coverage. In our simulation studies, these methods were effective in improving trial inferences when an external calibration dataset (containing information about error-prone and error-free measurements) with at least 15 subjects was available.\\
\\
To anticipate the impact of measurement error on trial inferences, the mechanism and magnitude of the measurement error should be considered. Endpoints that are measured with purely homoscedastic classical measurement error are expected to reduce the precision of treatment effect estimates and increase Type-II error at a given sample size, proportional to the relative amount of variance that is due to the error. Heteroscedastic classical error and differential error also affect Type-I error. Under systematic measurement error, only Type-I errors for testing null effects are expected to be at the nominal level. The treatment effect estimator itself is biased by systematic error and differential error. Heteroscedastic error can be addressed using standard robust standard error estimators (e.g. HC3 \cite{Long2000UsingModel}). Systematic error and differential error in the endpoint can be addressed via regression calibration.\\
\\
We considered regression calibration-like correction methods that rely on an external calibration set that contains information about both error-prone and error-free measurements. We anticipate such an external calibration set can be feasible as a planned pilot study phase of a trial. Our simulation study shows that the effectiveness of correction methods to adjust the trial results for endpoint measurement error are dependent on the size of the calibration sample and the strength of the correlation between the error-free and error-prone measurement of the trial endpoint. For a weak relation (R$^2$ = 0.20) we found the correction methods to be generally ineffective in improving trial inference with reasonably sized calibration sets (i.e., up to size N = 50). However, for medium (R$^2$ = 0.50) or strong (R$^2$ = 0.80) correlations, the regression calibration showed improvements with external calibration samples as small as $15$ observations. With the relatively small calibration samples (up to $50$ observations), our study showed that the Bootstrap method performed best in constructing confidence intervals in terms of coverage. The use of percentiles might explain that confidence intervals were slightly conservative (i.e. too broad) for small calibration samples ($10$ observations), and might be improved by using bias-corrected and accelerated Bootstrap intervals \cite{Hall1988}. The proposed calibration correction methods rely on a linear regression framework and can thus easily be extended to incorporate covariables in the trial analysis \cite{Senn1989CovariateTrials}.\\
\\
The use of measurement error corrections is still rare in applied biomedical studies despite an abundance of measurement error problems usually reported as an afterthought to a study \cite{Brakenhoff2018MeasurementReview,Shaw2018}. Indeed, to our knowledge, no measurement error correction methods have been used so far in the analysis of biomedical trials to correct for measurement error in the endpoint. This may in part be due to a common misconception that measurement error can only affect trial inference by reducing the precision of estimating the effect of treatment and increasing Type-II error, which can be improved by increasing the study sample size. Note that our study demonstrates that such an assumption is warranted only when strict classical homoscedastic error structure of the trial endpoint can be assumed. Such does not hold, for instance, when measurement errors are more pronounced in the tails of the distribution, or when measurement errors vary between treatment arms.\\
\\
Instead of the use of external calibration datasets, internal measurement correction approaches where both the preferred endpoint and the error contaminated endpoint are measured on a subset of trial participants may sometimes be more feasible. For internal calibration, Keogh et al. \cite{Keogh2016StatisticalTrials.} recently reviewed methods of moment estimation and maximum likelihood estimation approaches. There are also other approaches to correct for measurement error that we did not discuss in this paper. For instance, Cole and colleagues suggested a multiple imputing approach based on an internal calibration set \cite{Cole2006Multiple-imputationCorrection}. We also focused only on continuous outcomes in this paper. Problems and solutions for misclassified categorical outcomes can be found elsewhere \cite{Brooks2018TheResearch}. Yet, to the best of our knowledge, none of these methods have been tested in the setting where trial endpoints are measured with error and thus need further study.\\
\\
Lastly, we solely discuss parametric measurement error models, which might misspecify the measurement error model. The extent to which the distribution of the unmeasured outcome can be estimated without parametric assumptions is a question for further research. In the context of measurement error in explanatory variables this is formerly described as deconvolution (\cite{Carroll2006}, Chapter 12 and references therein).  Further, the method of non-parametric maximum likelihood has been successfully applied for explanatory variables measured with error \cite{Hesketh2003, Hesketh2001} and this might be an avenue of future research.\\
\\
In summary, the impact of measurement error in a continuous endpoint on trial inferences can be particularly non-ignorable when the measurement error is not strictly random, because Type-I error, Type-II and the effect estimates can be affected. To alleviate the detrimental effects of measurement error we proposed measurement error corrected estimators and a variety of methods to construct confidence intervals for non-random measurement error. To facilitate the implementation of these measurement error correction estimators we have developed the R package \texttt{mecor}, available at: www.github.com/LindaNab/mecor.

\newpage

\bibliographystyle{unsrt}
\bibliography{Mendeley.bib}

\newpage 

\begin{table}[ht]
\caption{Percentage bias, Empirical Standard Error (EmpSE), Squared root of Mean Squared Error (SqrtMSE), Coverage and average width of CI's of the naive estimator and the corrected estimator for systematic measurement error ($\theta_0 = 0$ and $\theta_1 = 1.05$ or $\theta_1 = 1.25$) for different values of R-squared and different sample sizes of the calibration data set. Each scenario is based on 10,000 replicates, the value of the estimand is 6.9, based on example trial 1 by Makrides et al. \cite{Makrides2003Efficacy13}}
\center{
\scalebox{0.6}{
\begin{tabular}{lllllcccccccc}
\textbf{Measure}$^*$& $\mathrm{R}^2_{Y^*,Y}$& $\theta_1$& & & \multicolumn{8}{c}{\textbf{Sample size external calibration set}} \\
 & & & Naive & & 5 & 7 & 10 & 15 & 20 & 30 & 40 & 50 \\ 
  \hline
Percentage bias (\%) & 0.8 & 1.25 & 24.9 &  & 88.9 & 29 & 3.7 & 2 & 1.6 & 0.9 & 0.7 & 0.4 \\ 
   & 0.8 & 1.05 & 4.9 &  & 88.9 & 29 & 3.7 & 2 & 1.6 & 0.9 & 0.7 & 0.4 \\ 
   & 0.5 & & 4.9 &  & 55.3 & 57.5 & -2.4 & 7.6 & 5.8 & 4.3 & 3 & 2 \\ 
   & 0.2 & & 4.9 &  & 168.2 & -62.6 & 98.8 & 33.4 & -142.2 & -28.3 & 23.9 & 14.6 \\ 
  EmpSE & 0.8 & 1.25 & 1.8 &  & 524.8 & 139.1 & 3 & 1.9 & 1.7 & 1.6 & 1.5 & 1.5 \\ 
   & 0.8 & 1.05 & 1.5 &  & 524.8 & 139.1 & 3 & 1.9 & 1.7 & 1.6 & 1.5 & 1.5 \\ 
   & 0.5 & & 1.9 &  & 267 & 329.1 & 83.7 & 14.4 & 11 & 2.5 & 2.3 & 2.1 \\ 
   & 0.2 & & 3 &  & 1131.2 & 210.8 & 723.2 & 462.2 & 1044.4 & 225.5 & 70.5 & 24.8 \\ 
  SqrtMSE & 0.8 & 1.25 & 2.5 &  & 524.8 & 139.1 & 3.1 & 1.9 & 1.7 & 1.6 & 1.5 & 1.5 \\ 
   & 0.8 & 1.05 & 1.5 &  & 524.8 & 139.1 & 3.1 & 1.9 & 1.7 & 1.6 & 1.5 & 1.5 \\ 
   & 0.5 & & 1.9 &  & 267 & 329.1 & 83.7 & 14.4 & 11 & 2.5 & 2.3 & 2.1 \\ 
   & 0.2 & & 3 &  & 1131.2 & 210.8 & 723.1 & 462.2 & 1044.4 & 225.5 & 70.5 & 24.8 \\ 
  Coverage (\%) & 0.8 & 1.25 & 83.5$^\ddagger$ & Zero-Variance & 70.3 & 74 & 77.4 & 80.3 & 82.8 & 84.4 & 85.3 & 86.3 \\ 
   &  &  &  & Delta & 93.8 & 95.3 & 95.7 & 95.9 & 96 & 96 & 95.9 & 95.7 \\ 
   &  &  &  & Fieller$^\dagger$ & - & - & 94.5 & 94.7 & 95 & 95.3 & 95.2 & 95 \\ 
   &  &  &  & Bootstrap & 95.9 & 96.1 & 95.5 & 94.9 & 94.8 & 95 & 95.1 & 94.8 \\ 
   & 0.8 & 1.05 & 94.6$^\ddagger$ & Zero-Variance & 77.8 & 81.3 & 84.4 & 87.1 & 89.2 & 90.9 & 92 & 92.2 \\ 
   &  &  &  & Delta & 92.1 & 93.9 & 94.3 & 94.8 & 95.1 & 95.3 & 95.4 & 95.2 \\ 
   &  &  &  & Fieller$^\dagger$ & - & - & 94.5 & 94.7 & 95 & 95.3 & 95.2 & 95 \\ 
   &  &  &  & Bootstrap & 95.9 & 96.1 & 95.5 & 94.9 & 94.8 & 95 & 95.1 & 94.8 \\ 
   & 0.5 & & 94.8$^\ddagger$ & Zero-Variance & 69.1 & 73.5 & 78.1 & 81.7 & 84.5 & 87.5 & 88.7 & 89.9 \\ 
   &  &  &  & Delta & 89.7 & 92 & 92.9 & 93.9 & 94.3 & 95.2 & 95.4 & 95.3 \\ 
   &  &  &  & Fieller$^\dagger$ & - & - & 94.5 & 95.2 & 95.2 & 95 & 94.8 & 94.9 \\ 
   &  &  &  & Bootstrap & 93.9 & 95.9 & 96.3 & 95.8 & 95.4 & 94.8 & 94.8 & 94.8 \\ 
   & 0.2 & & 95.1$^\ddagger$ & Zero-Variance & 57.1 & 64.5 & 71 & 76.8 & 80.3 & 84.3 & 86 & 87.6 \\ 
   &  &  &  & Delta & 86.8 & 89.7 & 90.9 & 92.2 & 93.5 & 94.4 & 94.6 & 94.9 \\ 
   &  &  &  & Fieller$^\dagger$ & - & - & 89.8 & 93.2 & 94.9 & 95.8 & 95.8 & 95.7 \\ 
   &  &  &  & Bootstrap & 88.9 & 93.8 & 95.5 & 96.4 & 96.7 & 96.8 & 96.8 & 96.1 \\ 
  Av. CI width & 0.8 & 1.25 & 6.9$^\ddagger$ & Zero-Variance & 30333 & 1141.5 & 5.5 & 4.7 & 4.7 & 4.6 & 4.5 & 4.5 \\ 
   &  &  &  & Delta & 40.7 & 13.6 & 8.7 & 7.5 & 7 & 6.5 & 6.3 & 6.1 \\ 
   &  &  &  & Fieller$^\dagger$ & - & - & 11.8 & 8.3 & 7 & 6.4 & 6.1 & 6 \\ 
   &  &  &  & Bootstrap & 86.9 & 29.3 & 14.1 & 8.3 & 7.1 & 6.4 & 6.1 & 6 \\ 
   & 0.8 & 1.05 & 5.8$^\ddagger$ & Zero-Variance & 36110.7 & 1359 & 6.5 & 5.6 & 5.5 & 5.4 & 5.4 & 5.4 \\ 
   &  &  &  & Delta & 35 & 12.2 & 8 & 7 & 6.7 & 6.3 & 6.1 & 6 \\ 
   &  &  &  & Fieller$^\dagger$ & - & - & 11.8 & 8.3 & 7 & 6.4 & 6.1 & 6 \\ 
   &  &  &  & Bootstrap & 86.9 & 29.3 & 14.1 & 8.3 & 7.1 & 6.4 & 6.1 & 6 \\ 
   & 0.5 &  & 7.4$^\ddagger$ & Zero-Variance & 7228.9 & 9759.5 & 763.1 & 37.5 & 17.8 & 7.7 & 7.3 & 7.1 \\ 
   &  &  &  & Delta & 58.1 & 43.2 & 21.2 & 12.6 & 11 & 9.3 & 8.7 & 8.4 \\ 
   &  &  &  & Fieller$^\dagger$ & - & - & 67.9 & 63.2 & 25 & 12.4 & 9.8 & 9 \\ 
   &  &  &  & Bootstrap & 146.8 & 87.4 & 65.2 & 34.7 & 22.8 & 12.4 & 9.9 & 9 \\ 
   & 0.2 &  & 11.6$^\ddagger$ & Zero-Variance & 126830.3 & 11677.5 & 87123.4 & 30709.4 & 324870.7 & 12430.8 & 774.6 & 126.8 \\ 
   &  &  &  & Delta & 179.3 & 102.5 & 112.7 & 69.9 & 65.7 & 34.1 & 19.7 & 16.6 \\ 
   &  &  &  & Fieller$^\dagger$ & - & - & 92.6 & 95.1 & 72.1 & 82.2 & 60.6 & 59.2 \\ 
   &  &  &  & Bootstrap & 176 & 121.9 & 126.2 & 118.7 & 107.7 & 77.6 & 54.8 & 39.7 \\ 
   \hline
\end{tabular}}}
\footnotesize{\\$^*$Monte Carlo standard errors of Bias, EmpSE, MSE and Coverage are subsequently, EmpSE$\sqrt{1/10,000}$; EmpSE$/(2\sqrt{9,999}$); $\sqrt{\frac{\sum_{i=1}^{10,000}[(\hat{\beta}_i-6.9)^2-\mathrm{MSE}]^2}{9,999\times10,000}}$ and $\sqrt{[\mathrm{Cover.}\times(1-\mathrm{Cover.})]/10,000}$ \cite{Morris2019}.\\
$^\dagger$Results of the Fieller method are shown if less than 5\% of cases resulted in undefined confidence intervals (see section \ref{sec:sim:comp}).\\
$^\ddagger$Coverage of the true intervention effect and average confidence interval width using regular Wald confidence intervals of the naive effect estimator.\\ 
Type-II error using the naive effect estimator is 0.2\%, 2.9\% and 31.6\% for $R^2_{Y^*,Y}=0.8$ (for both $\theta_1=1.05$ and $\theta_1=1.25$), $R^2_{Y^*,Y}=0.5$ and $R^2_{Y^*,Y}=0.2$, respectively. Type-II error using the corrected effect estimator using the Zero-Variance, Delta and Bootstrap method was 0\% in all scenario's. For the considered cases, Type-II error of the corrected effect estimator using the Fieller method was 0.2\% and 2.9\% for $R^2_{Y^*,Y}=0.8$ (for both $\theta_1=1.05$ and $\theta_1=1.25$) and $R^2_{Y^*,Y}=0.5$, respectively.}
\label{tbl:sme}
\end{table}

\begin{table}
\caption{Percentage bias, Emperical Standard Error (EmpSE), Mean Squared Error (MSE), Squared root of Mean Squared Error (SqrtMSE), Coverage and average width of CI's of the corrected estimator for differential measurement error ($\theta_{00} = 0$, $\theta_{10} = 1$, $\theta_{01} = 0$, $\theta_{11} = 1.05$) for the strong and poor proxy and different sample sizes of the calibration data set. Each scenario is based on 10,000 replicates, the value of the estimand is 6.9, based on example trial 1 by Makrides et al. \cite{Makrides2003Efficacy13}} 
\center{
\scalebox{0.65}{
\begin{tabular}{lllcccccc}
 \textbf{Measure$^*$}& $\mathrm{R}^2_{Y^*,Y}$ & & \multicolumn{5}{c}{\textbf{Sample size external calibration set}}\\
 & & Naive & & 10 & 20 & 30 & 40 & 50 \\ 
  \hline
Percentage bias (\%) & 0.8 & 91.8 &  & 5.2 & 1.2 & -0.4 & -0.2 & -0.1 \\ 
   & 0.5 & 91.8 &  & -9.7 & 33 & 154.2 & -21.4 & -0.1 \\ 
   & 0.2 & 91.9 &  & -319.4 & 152.9 & 193.1 & -21.5 & 2.2 \\ 
  EmpSE & 0.8 & 1.4 &  & 52 & 6.8 & 2.9 & 2.6 & 2.3 \\ 
   & 0.5 & 1.8 &  & 949.1 & 369.1 & 1080.4 & 142.1 & 4.5 \\ 
   & 0.2 & 2.9 &  & 2658 & 8425.8 & 1569.7 & 443.7 & 92.1 \\ 
  SqrtMSE & 0.8 & 6.5 &  & 52 & 6.8 & 2.9 & 2.6 & 2.3 \\ 
   & 0.5 & 6.6 &  & 949.1 & 369.1 & 1080.4 & 142.1 & 4.5 \\ 
   & 0.2 & 7 &  & 2658 & 8425.4 & 1569.7 & 443.7 & 92.1 \\ 
  Coverage (\%) & 0.8 & 0.7$^\ddagger$ & Zero-Variance & 43.8 & 59.9 & 67.9 & 72.7 & 76.8 \\ 
   &  &  & Delta & 97.1 & 96.6 & 96 & 95.7 & 95.9 \\ 
   &  &  & Bootstrap & 97.9 & 95.7 & 94.7 & 94.5 & 95 \\ 
   & 0.5 & 6.7$^\ddagger$ & Zero-Variance & 30.3 & 43.3 & 50.2 & 55.5 & 61 \\ 
   &  &  & Delta & 97.6 & 97.6 & 97.3 & 96.9 & 97 \\ 
   &  &  & Bootstrap & 98.4 & 98 & 96.6 & 95.8 & 95.5 \\ 
   & 0.2 & 41.1$^\ddagger$ & Zero-Variance & 25.7 & 35 & 41.9 & 46.6 & 52.2 \\ 
   &  &  & Delta & 98.4 & 99 & 98.9 & 98.9 & 98.9 \\ 
   &  &  & Bootstrap & 99 & 99.6 & 99.2 & 99 & 98.7 \\ 
  Av. CI width & 0.8 & 5.7$^\ddagger$ & Zero-Variance & 8.2 & 5.9 & 5.7 & 5.7 & 5.6 \\ 
   &  &  & Delta & 2688.7 & 18.3 & 12.1 & 10.5 & 9.5 \\ 
   &  &  & Bootstrap & 142.6 & 24.3 & 13.1 & 10.7 & 9.5 \\ 
   & 0.5 & 7.2$^\ddagger$ & Zero-Variance & 33 & 17.9 & 30.3 & 10.6 & 7.5 \\ 
   &  &  & Delta & 463975.1 & 49493.3 & 660587.5 & 13238 & 18.5 \\ 
   &  &  & Bootstrap & 303.5 & 118.8 & 58.4 & 34.2 & 24 \\ 
   & 0.2 & 11.4$^\ddagger$ & Zero-Variance & 64.6 & 150.5 & 53.1 & 43.1 & 26.8 \\ 
   &  &  & Delta & 1219162.5 & 26998502.1 & 486295.4 & 85139.8 & 3407.5 \\ 
   &  &  & Bootstrap & 562.9 & 353.8 & 283.3 & 221.4 & 170.2 \\ 
   \hline
\end{tabular}}}
\footnotesize{\\$^*$Monte Carlo standard errors of Bias, EmpSE, MSE and Coverage are subsequently, EmpSE$\sqrt{1/10,000}$; EmpSE/(2$\sqrt{9,999}$); $\sqrt{\frac{\sum_{i=1}^{10,000}[(\hat{\beta}_i-6.9)^2-\mathrm{MSE}]^2}{9,999\times10,000}}$ and $\sqrt{[\mathrm{Cover.}\times(1-\mathrm{Cover.})]/10,000}$ \cite{Morris2019}.\\
$^\ddagger$Coverage of the true intervention effect and average confidence interval width using regular Wald confidence intervals of the naive effect estimator.\\
Type-II error of the naive effect estimator was 0\%, 0\% and 0.4\% for $\mathrm{R}^2_{Y^*,Y}=0.8$, $\mathrm{R}^2_{Y^*,Y}=0.5$ and $\mathrm{R}^2_{Y^*,Y}=0.2$, respectively. Type-II error using the Zero-variance, Delta and Bootstrap method was 0\%.}
\label{tbl:dme}
\end{table}

\beginsupplement
\title{\vspace{-3cm} \centering Supplementary Materials}
\maketitle
\noindent These are the supplementary materials accompanying the paper `Measurement error in continuous endpoints in randomised trials: problems and solutions' by L. Nab et al. The supplementary materials are structured as follows. In section 1 we discuss two more example trials for illustration of measurement errors in an endpoint. In section 2 we explain why and under which assumptions ignoring measurement error will lead to incorrect inference. Section 3 provides an explanation of corrected effect estimators (and why these are consistent) and explains the methods used for confidence interval estimation. Section 4 provides a prove that measurement error depending on prognostic factors does not introduce bias in the treatment effect estimator.

\section{Illustrative examples}
We introduce here two additional example trials from literature, hypothesize that these trial could also have used endpoints measured with error to illustrate how the use of an endpoint that is contaminated with error would affect trial inference. We assume that the original endpoints used in our example trials are measurement error free.

\subsection{Example trial 2: energy expenditure}
Poehlman and colleagues \cite{Poehlman2002EffectsTrial} studied the effects of endurance and resistance training on total daily energy expenditure in a randomised trial of young sedentary women. Participants were randomized to one of three six-month during exercise programmes: endurance training, resistance training or the control arm. Some controversy regarding the effect of exercise training on total energy expenditure (TEE) existed at the time of the start of the trial, partly because of the difficulty to assess daily energy expenditure \cite{Poehlman2002EffectsTrial}. Starting 72 hours after completion of the training program, TEE of the participants was measured by doubly labelled water during a ten day period, which is considered the gold standard in measuring energy expenditure in humans \cite{Plasqui2007PhysicalWater}. In short, the study found no evidence for an effect of resistance and endurance training (compared to placebo) on total energy expenditure. Post-trial, measured TEE was higher in the control arm than in the two intervention arms. Table 1 shows the decrease in TEE of the women exposed to the existence training programme versus the placebo arm.

\subsection{Example trial 3: rheumatoid arthritis disease activity}
The U-Act-Early trial tested the efficacy of a new treatment strategy for rheumatoid arthritis (RA) in patients with newly diagnosed RA \cite{Bijlsma2017EarlyTrial} in a three-arm trial: tocilizumab plus methotrexate versus tocilizumab only versus methotrexate only, all as initial treatment. For endpoint assessment, this trial used a validated RA disease activity measure (the Disease Activity Score 28, DAS28) \cite{Prevoo1995ModifiedArthritis}) which is commonly used and recommended  to measure endpoints in RA clinical trials \cite{Pincus2009ComplexitiesMeasure, Anderson2012RheumatoidPractice}. In short, the trial showed that immediate initiation of tocilizumab with or without methotrexate is more effective than methotrexate alone to achieve sustained remission in newly diagnosed RA patients. The difference in mean DAS28 score in the tocilizumab plus methotrexate versus methotrexate only group after 24 weeks is shown in Table \ref{tab:resultstrials}. The sample size of the former groups reported in Table \ref{tab:resultstrials} is based on measurements available at 24 weeks of follow up.\\
\\
A common alternative approach to measure energy expenditure (example trial 2) is by a accelerometer, that measures body movement via motion sensors to assess energy expenditure (e.g. \cite{Plasqui2007PhysicalWater}). As compared to double labelled water (example trial 2), the accelerometer is cheaper, but less accurate \cite{Plasqui2007PhysicalWater}. Lastly, instead of endpoint assessment by DAS28 (example trial 3), where assessment is done by trained medical staff \cite{Prevoo1995ModifiedArthritis}, trials could alternatively use the patient-based RA disease activity score (PDAS), where endpoint assessment is done by the patient \cite{Choy2008DevelopmentPractice.}.\\
\\
For the example trial in the paper and each of the aforementioned example trials here, in Table \ref{tab:resultstrials} we show to what extent the Type-II of a test for treatment effect changes when a hypothetical lower standard of endpoint measurement would have been used introducing classical measurement error. The table clearly shows the anticipated increase in Type-II error with increasing error at the same sample size.\\
\begin{table}
\centering
\caption{Impact of classical measurement error on Type-II error}\label{tab:resultstrials} 
\begin{tabular}{cccccc}
& Effect estimate & Standard error & Sample Size & $\rho^\ddagger$ & Type-II error$^*$\\
\hline
Trial 1
& $6.9^\dagger$ & $1.27^\dagger$ &$393^\dagger$& $0$ & - \\
& & $2.43$ & $108$ & $0$ & $20\%$ \\
& & $2.71$ & $108$ &$1/5$ &$29\%$ \\
& & $2.45$ & $132$ &$1/5$ &$20\%$ \\
Trial 2 & $-246^\dagger$ & $369^\dagger$ &$35^\dagger$ & $0$ & -\\
& & $88.7$ & $600$ & 0 & $20\%$ \\
& & $109$ & $600$ & $1/3$ & $38\%$ \\
& & $88.7$ & $900$ & $1/3$ & $20\%$  \\
Trial 3 & $-1.4^\dagger$ & $0.08^\dagger$ & $198^\dagger$ & $0$ & - \\
& &$0.41$ &$8$ & 0 & $18\%$\\
& &$0.50$ &$8$ & $3/7$ & $41\%$ \\
& &$0.44$ &$12$ & $3/7$ & $18\%$ \\
\hline
\end{tabular}
\footnotesize{\\$\dagger$
Effect estimates, standard errors and sample sizes are based on results in papers by Makridis et al. \cite{Makrides2003Efficacy13} (trial 1), Poehlman et al. \cite{Poehlman2002EffectsTrial} (trial 2) and Bijlsma et al. \cite{Bijlsma2017EarlyTrial} (trial 3).\\
$\ddagger$ Proportion of observed variance in endpoints due to measurement error.\\
$*$ Type-II error calculations are based on results provided in section 3.1.}
\end{table}
\section{Measurement error structures}
Consider a two-arm randomized controlled trial that compares the effects of two treatments ($X \in \{0,1\}$), where 0 may represent a placebo treatment or an active comparator. Let $Y$ denote the true (or preferred) trial endpoint and $Y^*$ an error prone operationalisation of $Y$. We will assume that both $Y$ and $Y^*$ are measured on a continuous scale. Throughout, we assume that $Y^*$ is measured for all $i = 1,\hdots,N$ randomly allocated patients in the trial. We assume that the effect of allocated treatment ($X \in \{0,1\}$) on preferred endpoint $Y$ is defined by the linear model
\begin{eqnarray}\label{eq:sgen}
Y=\alpha_Y+\beta_YX+\varepsilon,
\end{eqnarray}
where $\beta_Y$ defines the treatment effect on the endpoint, and $\varepsilon$ has expected mean 0 and variance $\sigma^2$. Throughout, we assume that $X$ is fixed. Further, we assume that model \ref {eq:sgen} is inestimable from the observed data because the endpoint $Y^*$  instead of $Y$ was measured. We will assume that the relation between $Y$ and $Y^*$ is given by a linear model, 
\begin{equation}\label{eq:calmodel}
Y^*= \theta_0 + \theta_1Y + e,
\end{equation} 
where $e$ is a random variable whose distribution is independent of $\varepsilon$, $Y$ and $X$. The parameters $\theta_0$ and $\theta_1$ define the relation between $Y$ and $Y^*$, where it is assumed that $\theta_1$ does not equal 0. We assume that both parameters $\theta_0$ and $\theta_1$ are estimable only in the external calibration sample comprising individuals not included in the trial ($j = 1,\hdots,K$).\\
\\
Simple OLS regression estimators for $\beta_Y$, $\alpha_Y$ and $\sigma^2$ (the variance of the errors $\varepsilon$) in (\ref{eq:sgen}) are,
\begin{eqnarray}
\hat{\beta}_{Y^*}&=&\frac{\sum_i(X_i-\bar{X})(Y^*_i-\bar{Y^*})}{\sum_i(X_i-\bar{X})^2},\label{eq:beta_w}\\
\hat{\alpha}_{Y^*}&=&\bar{Y^*}-\hat{\beta}_{Y^*}\bar{X},\label{eq:alpha_w}\\
\omega_i&=&Y^*_i-\hat{\alpha}_{Y^*}-\hat{\beta}_{Y^*}X_i,\\\label{eq:omega_i} 
s^2&=&\frac{1}{N-2}\sum_i\omega_i^2,\label{eq:s2}
\end{eqnarray} 
respectively. In a two-arm trial, the interest is in making inferences about $\beta_Y$, which cannot be directly estimated because in the trial the endpoint of interest $Y$ was replaced by $Y^*$. In the following we will show: a) that $\hat{\beta}_{Y^*}$ may be a poor estimator for $\beta_Y$ (section 3.1-3.4), and b) how adjustments to $\hat{\beta}_{Y^*}$ using information from the calibration model described by ($\ref{eq:calmodel}$) can improve inference about the treatment effect (section 4). As a starting point, in the following section relevant and known properties are defined for the special case that $Y^*=Y$, which is then followed by the properties under different measurement error structures for $Y^*$ in subsequent sections. 

\subsection{No measurement error}
Consider the hypothetical case that $Y^*$ is a perfect proxy for $Y$, i.e. $Y^*=Y$. By using that $Y=\alpha_Y+\beta_YX+\varepsilon$, as defined in (\ref{eq:sgen}), it follows that:
\begin{equation*}
Y^*=\alpha_Y+\beta_YX+\varepsilon.
\end{equation*}
From standard regression theory (e.g. \cite{Davidson2004EconometricMethods}), we know that if the errors $\varepsilon$ satisfy the regular Gauss-Markov assumptions \cite{Davidson2004EconometricMethods} and their variance is defined by $\sigma^2$, the OLS estimators $\hat{\beta}_Y^*$, $\hat{\alpha}_Y^*$, and $s^2$ (defined by \ref{eq:beta_w}, \ref{eq:alpha_w}, and \ref{eq:s2}, respectively) are Best Linear Unbiased Estimators (BLUE) for $\beta_Y$, $\alpha_Y$, and $\sigma^2$, respectively.\\
\\
Moreover, if the $\varepsilon$ are independently and identically (iid) normally distributed, the OLS estimators $\hat{\beta}_{Y^*}$ and $\hat{\alpha}_{Y^*}$ (defined in \ref{eq:beta_w} and \ref{eq:alpha_w}, respectively) are the Maximum Likelihood Estimators (MLE) of $\beta_Y$ and $\alpha_Y$, respectively. Note that the errors $\varepsilon$ satisfy the Gauss-Markov assumptions if we assume that they are iid normally distributed with mean 0 and constant variance $\sigma^2$. 
\\
Hypotheses for the treatment effect $\beta_Y$, can be defined by
\begin{eqnarray*}
&H_0:& \beta_Y=\beta_0,\\
& H_A:& \beta_Y\neq\beta_0.
\end{eqnarray*}
Under normality  of the error terms $\varepsilon$, the OLS estimator $\hat{\beta}_Y^*$ defined in (\ref{eq:beta_w}) is the MLE for $\beta_Y$ and $s^2$ is an unbiased estimator for $\sigma^2$, the following is known for the Wald test:
\begin{equation}
T=\frac{\hat{\beta}_{Y^*}-\beta_0}{\sqrt{\widehat{\mathrm{Var}}(\hat{\beta}_{Y^*})}} \sim t_{N-2},\label{eq:ratio}
\end{equation}
where,
\begin{equation}
\widehat{\mathrm{Var}}(\hat{\beta}_{Y^*})=\frac{s^2}{\sum_i(X_i-\bar{X})^2}.\label{eq:varbeta}
\end{equation}
Assuming no measurement error in $Y$ and $X$, under $H_0$, $T$ follows a Student's t distribution with $N-2$ degrees of freedom \cite{Davidson2004EconometricMethods}. Under $H_A$, $T$ follows a Student's t distribution with $N-2$ degrees of freedom and non-centrality parameter $({\beta}_Y-\beta_0)/\sqrt{\widehat{\mathrm{Var}}(\hat{\beta}_{Y^*})}$.

\subsection{Classical measurement error} 
There is classical measurement error in $Y^*$ if $Y^*$ is an unbiased proxy for $Y$ \cite{Carroll2006}:
\begin{equation}
Y^*=Y+e, \label{eq:cme}
\end{equation}
where E$[e]=0$ and Var$(e)=\tau^2$ and $e$ mutually independent of $Y$, $X$, $\varepsilon$ (in (\ref{eq:sgen})).
\\
Using that $Y=\alpha_Y+\beta_YX+\varepsilon$ from (\ref{eq:sgen}), it follows that:
\begin{equation*}
Y^*=\alpha_Y+\beta_YX+\varepsilon+e.
\end{equation*}
Given the aforementioned assumptions, the sum of $e$ and $\varepsilon$, $ \delta_1 =e + \varepsilon$, has variance $\mathrm{Var}(\delta_1)= \sigma^2 +\tau^2$. It follows that if the errors $\delta_1$ satisfy the Gauss-Markov assumptions, $\hat{\beta}_{Y^*}$ in (\ref{eq:beta_w}) remains a BLUE estimator for $\beta_Y$. Also, $\hat{\alpha}_{Y^*}$ in (\ref{eq:alpha_w}) and $s^2$ in (\ref{eq:s2}) remain BLUE estimators for $\alpha_Y$ and the variance of $\delta_1$, respectively. \\
\\
Further, if $\delta_1$ is iid normally distributed with mean 0 and variance $\sigma^2+\tau^2$, then $\hat{\alpha}_{Y^*}$ is the MLE for $\alpha_Y$ and $\hat{\beta}_{Y^*}$ is the MLE for $\beta_Y$. Obviously, given that $\sigma^2>0$ and $\tau^2>0$, the variance of the OLS regression estimator $\hat{\beta}_{Y^*}$ is larger if there is classical measurement error in the outcome compared to the case when there is no measurement error. Under the null hypothesis, the Wald test-statistic $T$ defined in (\ref{eq:ratio}) still follows a Student's $t$ distribution with $N-2$ degrees of freedom. However, under the alternative hypothesis, the non-centrality parameter of $T$, $({\beta}_Y-\beta_0)/\sqrt{\widehat{\mathrm{Var}}(\hat{\beta}_{Y^*})}$, will be smaller in the presence of classical measurement error.\\
\\
To summarize, in the presence of only classical measurement error, Type-II error for detecting any given treatment effect increases, Type-I error is unaffected and the treatment effect estimator is unbiased MLE under standard regularity conditions.

\subsubsection{Heteroscedastic classical measurement error}

In the preceding we assumed that the Gauss-Markov assumptions were met. But notably, in the case that the variance of the errors $e$ in (\ref{eq:cme}) varies per treatment arm, the errors are no longer homoscedastic (as needed to satisfy the Gauss-Markov assumptions) but heteroscedastic. In the case of this type of heteroscedastic classical measurement error, it can be shown that the variance of $\beta_{Y^*}$ will be underestimated by the default estimator of the variance of $\hat{\beta}_{Y^*}$ defined by (\ref{eq:varbeta}), affecting both Type-I and Type-II error.

\subsection{Systematic measurement error}
There is systematic measurement error in $Y^*$, if $Y^*$ systematically depends on $Y$. Assuming this dependence is linear, the relation between $Y^*$ and $Y$ can be defined as:
\begin{equation}
Y^*=\theta_0+\theta_1Y+e,\label{eq:sme}
\end{equation}
where E$[e]=0$ and Var$(e)=\tau^2$. Throughout, we assume systematic measurement error if $\theta_0 \neq 0$  or $\theta_1 \neq 1$ (and of course, $\theta_1 \neq 0$  in all cases). We assume mutual independence between $e$ and $Y$, $X$, $\varepsilon$ ( in \ref{eq:sgen}). Naturally, if $\theta_0=0$ and $\theta_1=1$ the measurement error is of the classical form.\\
\\
By using that $Y=\alpha_Y+\beta_YX+\varepsilon$ from (\ref{eq:sgen}), it follows that:
\begin{equation*}
Y^*=\theta_0+\theta_1\alpha_Y+\theta_1\beta_YX+\theta_1\varepsilon + e.
\end{equation*}
Given the aforementioned assumptions, $\delta_2 = \theta_1\varepsilon + e$ with expected variance $\theta_1^2\sigma^2+\tau^2$. It follows that under the Gauss-Markov assumptions, $\hat{\beta}_{Y^*}$ defined in (\ref{eq:beta_w}) is BLUE for $\theta_1\beta_Y$, and $\hat{\alpha}_{Y^*}$ defined in (\ref{eq:alpha_w}) is BLUE for $\theta_0+\alpha_Y$ and $s^2$ defined in (\ref{eq:s2}) is BLUE for the variance of $\delta_2$ (i.e. $\theta_1^2\tau^2+\sigma^2$). Conversely, $\hat{\beta}_{Y^*}$ is no longer BLUE for $\beta_Y$. Note that in this case $s^2$ is BLUE for $\theta_1^2\sigma^2+\tau^2$, that is, depending on $\theta_1$, smaller or larger than $\sigma^2$ (the variance of the error terms if there is no measurement error).\\
\\
If we further assume that $\delta_2$ is iid normally distributed, we can conclude that $\hat{\alpha}_{Y^*}$ is the MLE for $\theta_0+\alpha_Y$ and $\hat{\beta}_{Y^*}$ is the MLE for $\theta_1\beta_Y$. Conversely, $\hat{\beta}_{Y^*}$ is no longer the MLE for $\beta_Y$, if there is systematic measurement error in $Y^*$. 
In the absence of a treatment effect, as $\theta_1\beta_Y=0$ if $\beta_Y=0$, $T$ defined in (\ref{eq:ratio}) still follows a Student's $t$ distribution with $N-2$ degrees of freedom. In the presence of any given treatment effect, $T$ follows a non-central Student's $t$ distribution with $N-2$ degrees of freedom and non-centrality parameter $(\theta_1\beta_Y-\beta_0)/\sqrt{\widehat{\mathrm{Var}}(\hat{\beta}_{Y^*})}$. Depending on the value of $\theta_1$, the non-centrality parameter will be smaller or larger than the non-centrality parameter in the absence of measurement error (see section 3.2).\\
\\
In summary, if there is systematic measurement error in the endpoints, the Type-I error is unaffected under standard regularity conditions and hence testing whether there is no effect is still valid under the null hypothesis \cite{Buonaccorsi1991MeasurementMeans}). Type-II, however, is affected (it may increase or decrease) and the treatment effect estimator is a biased MLE.

\subsection{Differential measurement error}
There is differential measurement error in $Y^*$ when measurement error varies with $X$. Assuming a linear model for this variation, formally:
\begin{equation}
Y^*=\theta_{00}+(\theta_{01}-\theta_{00})X+\theta_{10}Y+(\theta_{11}-\theta_{10})XY+ e_X,\label{eq:dme}
\end{equation}
where E$[e_X]=0$ and Var$(e_X)=\tau_X^2$ and $e_X$ independent of the endpoint of interest $Y$, and $\varepsilon$ in (\ref{eq:sgen}). From the equations it becomes clear that systematic error (equation (\ref{eq:sme})) can be seen as a special case of differential error, where $\theta_{00}=\theta_{01}$ and $\theta_{10}=\theta_{11}$.\\
\\
By using that $Y=\alpha_Y+\beta_YX+\varepsilon$ from (\ref{eq:sgen}), it follows from equation (\ref{eq:dme}) that,
\begin{equation*} 
Y^*=\theta_{00}+\theta_{10}\alpha_Y+\big[\theta_{01}-\theta_{00}+(\theta_{11}-\theta_{10})\alpha_Y+\theta_{11}\beta_Y\big]X+\big[\theta_{10} + (\theta_{11}-\theta_{10})X\big]\varepsilon+e_X.
\end{equation*}
Let $\delta_{3X} = \big[\theta_{10} + (\theta_{11}-\theta_{10})X\big]\varepsilon+e_X$, with expected variance $\big[\theta_{10}^2 + (\theta_{11}^2-\theta_{10}^2)X\big]\sigma^2+\tau_X^2$. Since the the error term $\delta_{3X}$ is no longer homoscedastic, the OLS estimators defined in (\ref{eq:beta_w}) and (\ref{eq:alpha_w}) are no longer BLUE. However, the OLS estimator $\hat{\beta}_{Y^*}$ in (\ref{eq:beta_w}) is consistent (although not efficient) for $\theta_{01}-\theta_{00}+(\theta_{11}-\theta_{10})\alpha_Y+\theta_{11}\beta_Y$. The OLS estimator $\hat{\alpha}_{Y^*}$ defined in (\ref{eq:alpha_w}) is consistent (although not efficient) for $\theta_{00}+\theta_{10}\alpha_Y$. Nevertheless, the estimator for the variance of $\hat{\beta}_{Y^*}$ defined in (\ref{eq:varbeta}) is no longer valid.\\
\\
By using the residuals $\omega_i$ defined in (\ref{eq:omega_i}), a heteroscedastic consistent estimator for the variance of $\hat{\beta}_{Y^*}$ is:
\begin{equation*}
\widehat{\mathrm{Var}}(\hat{\beta}_{Y^*})=\frac{\sum_i\left[{(X_i-\bar{X})^2\omega_i^2}\right]}{[\sum_i{(X_i-\bar{X})^2}]^2},
\end{equation*}
which is known as the White estimator \cite{Long2000UsingModel}. From standard regression theory, it is known that using the above defined estimator, $T$ defined in (\ref{eq:ratio}) is still valid. Yet, under differential measurement error no longer $\big[\theta_{01}-\theta_{00}+(\theta_{11}-\theta_{10})\alpha_Y+\theta_{11}\beta_Y\big]=0$ if $\beta_Y=0$. Thus, under the null hypothesis, $T$ defined in (\ref{eq:ratio}) follows a Student's $t$ distribution with $N-2$ degrees of freedom and non-centrality parameter $(\big[\theta_{01}-\theta_{00}+\theta_{11}\alpha_Y-\theta_{10}\alpha_Y+\theta_{11}\beta_0\big]-\beta_0)/\sqrt{\widehat{\mathrm{Var}}(\hat{\beta}_{Y^*})}$. Consequently, Type-I error changes if there is differential measurement error in $Y^*$ and test about contrast under the null hypothesis are invalid \cite{Buonaccorsi1991MeasurementMeans}. Moreover, under the alternative hypothesis, $T$ follows a non-central Student's $t$ distribution with $N-2$ degrees of freedom and non-centrality parameter $(\big[\theta_{01}-\theta_{00}+(\theta_{11}-\theta_{10})\alpha_Y+\theta_{11}\beta_Y\big]-\beta_0)/\sqrt{\widehat{\mathrm{Var}}(\hat{\beta}_{Y^*})}$. Depending on the values of the $\theta$'s and $\alpha_Y$, the non-centrality parameters will be smaller or larger than 0 and the non-centrality parameter if there is no measurement error, respectively (see section 3.2). Hence, Type-I error and Type-II error could increase or decrease if there is differential measurement error in $Y^*$.\\
\\
To summarize, Type-I error is not expected nominal ($\alpha$) if there is differential measurement error in $Y^*$ (see also \cite{Buonaccorsi1991MeasurementMeans}). Also, similar to systematic error in $Y^*$, Type-II error is affected (may increase or decrease) and the treatment effect estimator is a biased estimator. 

\section{Correction methods for measurement error in continuous endpoints}
To accommodate measurement error correction, we assume that $Y$ and $Y^*$ are both measured for a smaller set of different individuals not included in the trial ($j = 1,\hdots,K, K < N$), hereinafter referred to as the external calibration sample. In all but one case, it is assumed that only $Y^*$ and $Y$ are measured in the external calibration sample. In the case that the error in $Y^*$ is different for the two treatment groups, it is assumed that the external calibration sample is in the form of a small pilot study where both treatments are allocated (i.e., $Y^*$ and $Y$ are both measured after assignment of $X$).

\subsection{Systematic measurement error}
Using an external calibration set and assuming that the errors $e$ in (\ref{eq:sme}) are iid normal, the MLE of the measurement error parameters in (\ref{eq:sme}) are:
\begin{eqnarray}
\hat{\theta}_{1}&=&\frac{\sum_j(Y_j^{(c)}-\bar{Y}^{(c)})(Y^{*(c)}_j-\bar{Y}^{*(c)})}{\sum(Y_j^{(c)}-\bar{Y}^{(c)})^2},\label{eq:theta1sme}\\
\hat{\theta}_{0}&=&\bar{Y}^{*(c)}-\hat{\theta}_{1}\bar{Y}^{(c)},\nonumber\\
t^2&=&\frac{1}{K-2}\sum_j (Y^{*(c)}_j-\hat{\theta}_{0}-\hat{\theta}_{1}Y_j^{(c)})^2.\nonumber
\end{eqnarray}
The superscript (c) is used to indicate that the measurement is obtained in the calibration set. From section 3.4, under systematic measurement error and assuming that $\varepsilon$ in (\ref{eq:sgen}) and $e$ in (\ref{eq:sme}) iid normal and independent, the estimator $\hat{\beta}_{Y^*}$ defined in (\ref{eq:beta_w}) is the MLE of $\theta_1\beta_Y$ and, the estimator $\hat{\alpha}_{Y^*}$ defined in (\ref{eq:alpha_w}) is the MLE of $\theta_0+\theta_1\alpha_Y$. Natural sample estimators for $\alpha_Y$ and $\beta_Y$ are then
\begin{align}\label{eq:sestsme}
\hat{\alpha}_Y=(\hat{\alpha}_{Y^*}-\hat{\theta}_0)/\hat{\theta}_1  \text{\quad and \quad}
\hat{\beta}_Y=\hat{\beta}_{Y^*}/\hat{\theta}_1,
\end{align}
where $\hat{\theta}_0$ and $\hat{\theta}_1$ are the estimated error parameters from the calibration data set. From equation (\ref{eq:sestsme}), it becomes apparent that $\hat{\theta}_1$ needs to be assumed bounded away from zero for finite estimates of  $\hat{\alpha}_Y$  and $\hat{\beta}_Y $ \cite{Buonaccorsi2010MeasurementApplications}.
\\
The first moment of estimators $\hat{\alpha}_Y$ and $\hat{\beta}_Y$ can be approximated by using multivariate Taylor expansions and assuming that ($\hat{\alpha}_{Y^*}$, $\hat{\beta}_{Y^*}$, $\hat{\theta}_0$, $\hat{\theta}_1$) are normally distributed \cite{Buonaccorsi2010MeasurementApplications}, 
\begin{align*}
\mathrm{E}[\hat{\alpha}_Y]\approx \alpha_Y+\frac{\big[\alpha_Y-\bar{y^*}\big]\tau^2}{\theta_1^2S_{yy}^{(c)}} \text{\quad and \quad}
\mathrm{E}[\hat{\beta}_Y]\approx \beta_Y+\frac{\beta_Y\tau^2}{\theta_1^2S_{yy}^{(c)}},
\end{align*}
where $S_{yy}^{(c)}=\sum(Y_j^{(c)}-\bar{Y}^{(c)})^2$, the total sum of squares of $Y^{(c)}$. In conclusion, the estimators $\hat{\alpha}_Y$ and $\hat{\beta}_Y$ are consistent. Formal derivations for the presented formulas are provided in the Appendix.\\
\\
In the following we will focus on specifying confidence limits for the treatment effect estimator $\hat{\beta}_Y$ defined in (\ref{eq:sestsme}). We make use of the fact that this estimator is a ratio, which motivates the use of the Delta method, Fieller method and Zero-variance method \cite{Franz2007Ratios:Use}. We also present a non-parametric bootstrap method for specifying confidence limits \cite{Efron1979BootstrapJackknife}.

\subsubsection{Delta method} 
Assuming that $\hat{\beta}_{Y^*}$ and $\hat{\theta}_1$ are both normally distributed and applying the Delta method, the second moment of $\hat{\beta}_Y$ can be approximated \cite{Buonaccorsi1991MeasurementMeans}. Formal derivations of the presented formulas are provided in Appendix A. The Delta method variance of $\hat{\beta}_Y$ is given by:
\begin{align*}
\mathrm{Var}\big(\hat{\beta}_Y\big)&\approx\frac{1}{\theta_1^2}\Big[\frac{\theta_1^2\sigma^2+\tau^2}{S_{xx}}+\frac{\beta_Y^2\tau^2}{S_{yy}^{(c)}}\Big],
\end{align*}
where $S_{xx}=\sum_i(X_i-\bar{X})^2$, the total sum of squares of $X$. An approximation of the above defined variance, denoted by $\widehat{\text{Var}}(\hat{\beta}_Y)$, is provided by approximating $\theta_1$, $\theta_1^2\sigma^2+\tau^2$, $\tau^2$ and $\beta_Y$ respectively by $\hat{\theta}_1$, $s^2$, $t^2$ and $\hat{\beta}_Y$ \cite{Buonaccorsi1991MeasurementMeans}.\\
\\
An approximate confidence interval for the estimator $\hat{\beta}_Y$ is then given by
\begin{equation}\label{eq:cidelta}
\hat{\beta}_Y \pm t_{(\alpha/2,n-2)}\sqrt{\widehat{\mathrm{Var}}\big(\hat{\beta}_Y\big)}.
\end{equation}

\subsubsection{Fieller method}
A second method to construct confidence intervals for the estimator $\hat{\beta}_Y$ in (\ref{eq:sestsme}), described by Buonaccorsi, is the Fieller method \cite{Buonaccorsi1991MeasurementMeans,Fieller1940TheInsulin}. In the case that $\hat{\theta}_1$ is significantly different from zero at a significance level of $\alpha$ (that is, $\hat{\theta}_1/\sqrt{t^2/S^{(c)}_{yy}}>t_{N-2}$), the $(1-\alpha)$ confidence intervals of $\hat{\beta}_Y$ are defined by the Fieller method by:
\begin{equation}\label{eq:cifieller}
l_{upper,lower}=\frac{\hat{\beta}_{Y^*}\hat{\theta}_1 \pm \sqrt{\hat{\beta}^2_{Y^*}\hat{\theta}_1^2-(\frac{t^2}{S^{(c)}_{yy}}t_q^2-\hat{\theta}_1^2)(\frac{s^2}{S_{xx}}t_q^2-\hat{\beta}_{Y^*}^2)}}{\frac{\tau^2}{S^{(c)}_{yy}}t_q^2+\hat{\theta}_1^2}.
\end{equation} 
A formal derivation can be found in Appendix A. 

\subsubsection{Zero-variance method} 
The zero-variance method adjusts the observed endpoints $Y^*_i$ by
\begin{equation*}\label{eq:adjval}
\hat{Y_i }=(Y^*_i-\hat{\theta}_0)/\hat{\theta}_1,
\end{equation*}
where $\hat{\theta}_0$ and $\hat{\theta}_1$ are derived from (\ref{eq:sme}). The adjusted endpoints are regressed on the treatment variable $X$, which yields,
\begin{align*}\label{eq:naive}
\hat{\beta}_{\hat{Y}}&=\frac{\sum_i(X_i-\bar{X})(\hat{Y}_i-\bar{\hat{Y}})}{\sum_i(X_i-\bar{X})^2}=\frac{\sum_i(X_i-\bar{X})(Y^*_i-\bar{Y^*})/\hat{\theta}_1}{\sum_i(X_i-\bar{X})^2}=\hat{\beta}_{Y^*}/\hat{\theta}_1,\\
\hat{\alpha}_{\hat{Y}}&=\bar{\hat{Y}}-\hat{\beta}_{\hat{Y}}\bar{X}=\frac{\bar{Y^*}-\hat{\beta}_{Y^*}\bar{X}-\hat{\theta}_0}{\hat{\theta}_1}=(\hat{\alpha}_{Y^*}-\hat{\theta}_0)/\hat{\theta}_1,\\ 
s_{\hat{Y}}^2&=\frac{1}{N-2}\sum_i(\hat{Y}_i-\hat{\alpha}_{\hat{Y}}-\hat{\beta}_{\hat{Y}}X_i )^2=\frac{1}{\hat{\theta}_1^2}s^2,
\end{align*} 
with $\hat{\beta}_{Y^*}$, $\hat{\alpha}_{Y^*}$ and $s^2$ as in equations (\ref{eq:beta_w}, \ref{eq:alpha_w} and \ref{eq:s2}), respectively. Thus, $\hat{\beta}_{\hat{Y}}$ equals $\hat{\beta}_Y$ and $\hat{\alpha}_{\hat{Y}}$ equals  $\hat{\alpha}_Y$ defined in (\ref{eq:sestsme}).\\
\\
If the value of $\hat{\theta}_1$ (i.e. $\theta_1$) is known, the variance of the estimator $\hat{\beta}_{\hat{Y}}$ is equal to:
\begin{equation*}
\text{Var}(\hat{\beta}_{\hat{Y}})=\text{Var}(\hat{\beta}_{Y^*})/\theta_1^2=\frac{\sigma^2+\tau^2/\theta_1^2}{\sum_i{(X_i-\bar{X})^2}}.
\end{equation*}
Using the standard OLS regression framework the variance of $\hat{\beta}_{\hat{Y}}$ can be estimated by:
\begin{equation}\label{eq:zv}
\widehat{\text{Var}}(\hat{\beta}_{\hat{Y}})=\frac{s_{\hat{Y}}^2}{\sum_i(X_i-\bar{X})^2}=\frac{s^2/\hat{\theta}_1^2}{\sum_i(X_i-\bar{X})^2}.
\end{equation}
By replacing $\hat{\theta}_1$ by $\theta_1$ in the above, the quantity in (\ref{eq:zv}) is in expectation equal to $\text{Var}(\hat{\beta}_{\hat{Y}})$ (defined above). The quantity in (\ref{eq:zv}) is used in the zero-variance method to construct confidence intervals for $\hat{\beta}_{\hat{Y}}$, by replacing $\widehat{\text{Var}}(\hat{\beta}_{\hat{Y}})$ for $\widehat{\text{Var}}(\hat{\beta}_{Y})$ in equation \ref{eq:cidelta}. In conclusion, this zero-variance approach will provide confidence intervals for the treatment effect estimator while assuming there is no variance in $\hat{\theta}_1$ (giving it its name zero-variance method). Although the zero-variance approach wins in terms of simplicity, it may underestimate the variability of the ratio since the variance in $\hat{\theta}_1$ is assumed zero.

\subsubsection{Bootstrap}
An alternative for defining confidence intervals for the corrected treatment effect estimator $\hat{\beta_Y}$ is by using a non-parametric bootstrap \cite{Efron1979BootstrapJackknife}. We propose the following stepwise procedure:
\begin{enumerate}
\item Draw a random sample with replacement of size $K$ of the calibration sample $(Y^{*(c)},Y^{(c)})$ to estimate $\hat{\theta}_{1_B}$ defined in (\ref{eq:theta1sme}).
\item Draw a random sample with replacement of size $N$ of the trial data $(Y^*,X)$ to calculate the corrected treatment effect estimate by $\hat{\beta}_{Y_B}=\beta_{Y^*_B}/\hat{\theta}_{1_B}$. Where $\beta_{Y^*_B}$ is defined in (\ref{eq:beta_w}).
\item Repeat step 1-2 $B$ times, with $B$ large (e.g. 999 times). 
\item Approximate confidence intervals are given by the $(\alpha/2,1-\alpha/2)$ percentile of the distribution of $\hat{\beta}_{Y_B}$.
\end{enumerate}

\subsection{Differential measurement error}
For corrections for endpoints that suffer from differential measurement error we will here assume the existence of a pilot trial, which serves as an external calibration set, where both treatments are allocated at random that serves as an external calibration set to estimate the measurement error model in (\ref{eq:dme}). For notational convenience we rewrite the linear model in equation (\ref{eq:dme}) in matrix form as:
\begin{eqnarray}\label{eq:dmematrix}
\boldsymbol{Y^*}=\boldsymbol{X}\boldsymbol{\theta}+\boldsymbol{e},
\end{eqnarray}
where E$(\boldsymbol{e}) = 0$ and E$(\boldsymbol{ee}')=\boldsymbol{\Sigma}$, a positive definite matrix, with $\tau^2_X$ on its diagonal. Further, $\boldsymbol{\theta}=(\theta_1, \theta_2, \theta_3, \theta_4)=(\theta_{00}, \theta_{01} - \theta_{00}, \theta_{10}, \theta_{11}-\theta_{10})$. In the external calibration set, the measurement error parameters $\boldsymbol{\hat{\theta}}$ can be estimated by,
\begin{equation}\label{eq:esttheta}
\boldsymbol{\hat{\theta}}=(\boldsymbol{X^{(c)}}'\boldsymbol{X^{(c)}})^{-1}\boldsymbol{X^{(c)}}'\boldsymbol{Y^{(c)}},
\end{equation}
with variance,
\begin{equation*}
\text{Var}(\boldsymbol{\hat{\theta}}) = (\boldsymbol{X^{(c)}}'\boldsymbol{X^{(c)}})^{-1}\boldsymbol{X^{(c)}}'\boldsymbol{\Sigma} \boldsymbol{X^{(c)}}(\boldsymbol{X^{(c)}}'\boldsymbol{X^{(c)}})^{-1}.
\end{equation*}
See \cite{Long2000UsingModel} for a discussion on different estimators for the above defined variance. From section 2.5 it follows that natural estimators for $\alpha_Y$ and $\beta_Y$ are, 
\begin{equation}\label{eq:sestdme}
\hat{\alpha}_Y=(\hat{\alpha}_{Y^*}-\hat{\theta}_{00})/\hat{\theta}_{10} \textrm{\quad and \quad}
\hat{\beta}_Y=(\hat{\beta}_{Y^*}+\hat{\alpha}_{Y^*}-\hat{\theta}_{01})/\hat{\theta}_{11}-\hat{\alpha}_{Y},
\end{equation}
where $\hat{\theta}_{00}$, $\hat{\theta}_{10}$, $\hat{\theta}_{01}$ and $\hat{\theta}_{11}$ are estimated from the external calibration set. Here it is assumed that both $\hat{\theta}_{10}$ and $\hat{\theta}_{11}$ are bounded away from zero (for reasons similar to those mentioned in section 3.1).\\
\\
By multivariate Taylor expansions, the first moments of the estimators $\hat{\alpha}_Y$ and $\hat{\beta}_Y$ defined in (\ref{eq:sestdme}) can be approximated \cite{Buonaccorsi1991MeasurementMeans}, in the same way as the estimators for systematic measurement error (section 4.1),
\begin{flalign*}
\mathrm{E}\big[\hat{\alpha}_{Y}\big]&\approx
\alpha_Y+\frac{1}{\theta_{10}^2}\Big[\alpha_Y\text{Var}\big(\hat{\theta}_{10}\big) + \text{Cov}\big(\hat{\theta}_{00},\hat{\theta}_{10}\big)\Big],\\
\mathrm{E}\big[\hat{\beta}_{Y}\big]&\approx\beta_Y+\frac{1}{\theta_{11}^2}\Big[(\beta_Y + \alpha_Y)\text{Var}\big(\hat{\theta}_{11}\big)+\text{Cov}\big(\hat{\theta}_{01}, \hat{\theta}_{11}\big)\Big]\\&-\frac{1}{\theta_{10}^2}\Big[\alpha_Y\text{Var}\big(\hat{\theta}_{10}\big)+\text{Cov}\big(\hat{\theta}_{00}, \hat{\theta}_{10}\big)\Big]. 
\end{flalign*}
From this, it is apparent that the estimators $\hat{\alpha}_Y$ and $\hat{\beta}_Y$ defined in (\ref{eq:sestdme}) are consistent (details are found in the Appendix). In the subsequent sections we review the Delta method, zero-variance and propose a bootstrap for specifying confidence limits for the estimator of the treatment effect under differential measurement error of the endpoints.

\subsubsection{Delta method}
The variance of the estimator $\hat{\beta}_Y$ defined in (\ref{eq:sestdme}) can be approximated by the Delta method \cite{Buonaccorsi1991MeasurementMeans}:
\begin{align*}
\mathrm{Var}\big(\hat{\beta}_Y\big)&\approx\frac{1}{\theta_{11}^2}\Big[\big(\beta_Y+\alpha_Y\big)^2\text{Var}\big(\hat{\theta}_{11}\big)+\text{Var}\big(\hat{\beta}_{Y^*}\big)+\text{Var}\big(\hat{\alpha}_{Y^*}\big)+\\&2\text{Cov}\big(\hat{\alpha}_{Y^*},\hat{\beta}_{Y^*}\big)+\text{Var}\big(\hat{\theta}_{01}\big)+2\big(\beta_Y+\alpha_Y\big)\text{Cov}\big(\hat{\theta}_{11},\hat{\theta}_{01}\big)\Big]+\\&\mathrm{Var}\big(\hat{\alpha}_Y\big),
\end{align*}
where $\mathrm{Var}\big(\hat{\alpha}_Y\big)$ is approximated by:
\begin{align*}
\mathrm{Var}\big(\frac{\hat{\alpha}_{Y^*}-\hat{\theta}_{00}}{\hat{\theta}_{10}}\big)&\approx\
\frac{1}{\theta_{10}^2}\Big[\text{Var}\big(\hat{\alpha}_{Y^*}\big)+\alpha_Y^2\text{Var}\big(\hat{\theta}_{10}\big)+\text{Var}\big(\hat{\theta}_{0
0}\big)+2\alpha_Y\text{Cov}\big(\hat{\theta}_{00},\hat{\theta}_{10}\big)\Big].
\end{align*}
An approximate confidence interval for the estimator $\hat{\beta}_Y$ in (\ref{eq:sestdme}) is:\\
\begin{equation}\label{eq:cideltadme}
\hat{\beta}_Y \pm t_{(\alpha/2,n-2)}\sqrt{\mathrm{Var}\big(\hat{\beta}_Y\big)}.
\end{equation}
An approximation of $\theta_{11}$, $\theta_{10}$, $\theta_{11}^2\sigma^2+\tau_1^2$, $\theta_{10}^2\sigma^2+\tau_0^2$, $\tau_1^2$, $\tau_0^2$, $\beta_Y$ and $\alpha_Y$ in the above is provided by: $\hat{\theta}_{11}$, $\hat{\theta}_{10}$, $s_1^2$, $s_0^2$, $t_1^2$, $t_0^2$, $\hat{\beta}_Y$ and $\hat{\alpha}_Y$ \cite{Buonaccorsi1991MeasurementMeans}.   

\subsubsection{Zero-variance method}
The zero-variance method adjusts the observed endpoints $Y^*_i$ by
\begin{equation*}
\hat{Y}_{ix}=(Y_{ix}^*-\hat{\theta}_{0x})/\hat{\theta}_{1x}, 
\end{equation*}
for $x \in \{0,1\}$ and $\hat{\theta}_{0x})$ and $\hat{\theta}_{1x}$ derived from (\ref{eq:esttheta}). In the zero-variance method the above defined adjusted values are regressed on the treatment variable $X$, yielding in estimators $\hat{\alpha}_{\hat{Y}}$ and $\hat{\beta}_{\hat{Y}}$, which are, respectively, equal to the estimators $\hat{\alpha}_Y$ and $\hat{\beta}_Y$ defined in (\ref{eq:sestdme}). The variance of these estimators can be approximated with a heteroscedastic consistent covariance estimator (see \cite{Long2000UsingModel} for an overview). Confidence intervals for $\hat{\beta}_{\hat{Y}}$ are subsequently constructed by using formula \ref{eq:cideltadme}. Similar to what is described in section 4.1.3 discussing the zero-variance method for systematic measurement error, this way of constructing confidence intervals neglects the variance of the $\theta$'s from the calibration data set, and will thus often yield in confidence intervals that are too narrow.

\subsubsection{Bootstrap}
We here alternatively propose a non-parametric bootstrap procedure to specify confidence limits. This entails the following steps:
\begin{enumerate}
\item Draw a random sample with replacement of size $K$ of the calibration sample and estimate $\boldsymbol{\hat{\theta}}$ as defined in (\ref{eq:esttheta}).
\item Draw a random sample (with replacement) of size $N$ of the study population and calculate the effect estimate by $\hat{\alpha}_{Y_B}=(\alpha_{Y^*_B}-\hat{\theta}_{{00}_B})/\hat{\theta}_{{10}_B}$ and $\hat{\beta}_{Y_B}=(\beta_{Y^*_B}+\alpha_{Y^*_B}-\hat{\theta}_{{01}_B})/\hat{\theta}_{{11}_B}-\hat{\alpha}_{Y_B}$. Where $\beta_{Y^*_B}$ and $\alpha_{Y^*_B}$ are defined in (\ref{eq:beta_w}) and (\ref{eq:alpha_w}), respectively.
\item Repeat step 1-2 $B$ times, with $B$ large (e.g. 999 times). 
\item Approximate confidence intervals are given by the $(\alpha/2,1-\alpha/2)$ percentile of the distribution of $\hat{\beta}_{Y_B}$.
\end{enumerate}

\section{Measurement error depending on prognostic factors}
Suppose that there is a prognostic factor $S$, and assume that, $\mathrm{E}[Y|X,S] = \alpha_Y + \beta_Y X + \gamma_Y S$, $\mathrm{E}[Y^*|Y,S]= Y + \zeta S$, $Y^* \indep X|Y$ (non-differential measurement error) and $S \indep X$ (randomization is well-performed).\\
\\
Suppose we want to estimate the effect of $Y$ on $X$ (i.e., $\beta_Y$), but instead of $Y$ we have only measured the with measurement error contaminated $Y^*$. If one is aware that there is a prognostic factor that confounds the relation between $Y^*$ and $Y$ (and this factor is measured), one could decide to regress $Y^*$ on $X$ and $S$. The regression of $Y^*$ on $X$ and $S$ equals, 
\begin{eqnarray*}
\mathrm{E}[Y^*|X,S]&=&\mathrm{E}_{Y|X,S}\{\mathrm{E_{Y^*|X,S,Y}[Y^*|X,S,Y]|X,S}\}\\
&=& \mathrm{E}_{Y|X,S}\{\mathrm{E_{Y^*|S,Y}[Y^*|S,Y]|X,S}\}\\
&=& \mathrm{E}_{Y|X,S}\{\mathrm{Y+\zeta S|X,S}\}\\
&=& \alpha_Y + \beta_Y X + (\gamma_Y + \zeta)S.
\end{eqnarray*}
Thus, using the with measurement error contaminated endpoint $Y^*$ instead of the preferred endpoint $Y$ will provide an unbiased estimation of $\beta_Y$.\\
\\
However, if one is not aware of the prognostic factor, one might naively regress $Y^*$ on $X$, which equals:
\begin{eqnarray*}
\mathrm{E}[Y^*|X]&=&\mathrm{E}_{S|X}\{\mathrm{E}_{Y|X,S}\{\mathrm{E_{Y^*|X,S,Y}[Y^*|X,S,Y]|X,S}\}|X\}\\
&=& \mathrm{E}_{S|X}\{\alpha_Y + \beta_Y X + (\gamma_Y + \zeta)S|X\}\\
&=& \alpha_Y + \beta_Y X + (\gamma_Y + \zeta)\mathrm{E}[S].
\end{eqnarray*}
In conclusion, by ignoring the prognostic factor and using the with measurement error contaminated endpoint $Y^*$ instead of the preferred endpoint $Y$, the regression of $Y^*$ on $X$ still results in an unbiased estimation of $\beta_Y$.

\newpage
\appendix
\renewcommand{\thesection}{A\arabic{section}}
\renewcommand{\thesubsection}{A\arabic{section}.\arabic{subsection}}
\renewcommand{\theequation}{A\arabic{equation}}
\setcounter{equation}{0}
\section{Approximation of bias and variance in corrected estimator}\label{appendix}
\subsection{Systematic measurement error}
Obvious estimators for $\alpha_Y$ and $\beta_Y$ are:
\begin{align*}
\hat{\alpha}_Y=(\hat{\alpha}_{Y^*}-\hat{\theta}_0)/\hat{\theta}_1 &\textrm{\quad and \quad}
\hat{\beta}_Y=\hat{\beta}_{Y^*}/\hat{\theta}_1.
\end{align*}
These estimators can be approximated with a second order Taylor expansion by:
\begin{align*}
\frac{\hat{\beta}_{Y^*}}{\hat{\theta}_1}&\approx\frac{\beta_{Y^*}}{\theta_1}-\frac{\beta_{Y^*}}{\theta_1^2}(\hat{\theta}_1-\theta_1)+\frac{1}{\theta_1}(\hat{\beta}_{Y^*}-\beta_{Y^*})\\
&+\frac{1}{2!}\Big[ \frac{2\beta_{Y^*}}{\theta_1^3}(\hat{\theta}_1-\theta_1)^2- \frac{2}{\theta_1^2}(\hat{\theta}_1-\theta_1)(\hat{\beta}_{Y^*}-\beta_{Y^*})\Big],\\
\frac{\hat{\alpha}_{Y^*}}{\hat{\theta}_1}&\approx\frac{\alpha_{Y^*}}{\theta_1}-\frac{\alpha_{Y^*}}{\theta_1^2}(\hat{\theta}_1-\theta_1)+\frac{1}{\theta_1}(\hat{\alpha}_{Y^*}-\alpha_{Y^*})\\
&+\frac{1}{2!}\Big[ \frac{2\alpha_{Y^*}}{\theta_1^3}(\hat{\theta}_1-\theta_1)^2- \frac{2}{\theta_1^2}(\hat{\theta}_1-\theta_1)(\hat{\alpha}_{Y^*}-\alpha_{Y^*})\Big],\\
\frac{\hat{\theta}_0}{\hat{\theta}_1}&\approx\frac{\theta_0}{\theta_1}-\frac{\theta_0}{\theta_1^2}(\hat{\theta}_1-\theta_1)+\frac{1}{\theta_1}(\hat{\theta}_0-\theta_0)\\
&+\frac{1}{2!}\Big[ \frac{2\theta_0}{\theta_1^3}(\hat{\theta}_1-\theta_1)^2- \frac{2}{\theta_1^2}(\hat{\theta}_1-\theta_1)(\hat{\theta}_0-\theta_0)\Big].
\end{align*}
Simplifying these terms and substraction of the latter two, will lead to the following approximations for $\hat{\alpha}_Y$ and $\hat{\beta}_Y$:\\
\begin{align*}
\frac{\hat{\beta}_{Y^*}}{\hat{\theta}_1}&\approx \frac{\beta_{Y^*}}{\theta_1}+\frac{1}{\theta_1}\Big[-\frac{\beta_{Y^*}}{\theta_1}(\hat{\theta}_1-\theta_1)+(\hat{\beta}_{Y^*}-\beta_{Y^*})\Big]\\
&+\frac{1}{\theta_1^2}\Big[\frac{\beta_{Y^*}}{\theta_1}(\hat{\theta}_1-\theta_1)^2-(\hat{\beta}_{Y^*}-\beta_{Y^*})(\hat{\theta}_1-\theta_1)\Big],\\
\frac{\hat{\alpha}_{Y^*}-\hat{\theta}_0}{\hat{\theta}_1}&\approx\frac{\alpha_{Y^*}-\theta_0}{\theta_1}+\frac{1}{\theta_1}\Big[-\frac{\alpha_{Y^*}-\theta_0}{\theta_1}(\hat{\theta}_1-\theta_1)+(\hat{\alpha}_{Y^*}-\alpha_{Y^*})-(\hat{\theta}_0-\theta_0)\Big]\\
&+\frac{1}{\theta_1^2}\Big[ \frac{\alpha_{Y^*}-\theta_0}{\theta_1}(\hat{\theta}_1-\theta_1)^2-(\hat{\alpha}_{Y^*}-\alpha_{Y^*})(\hat{\theta}_1-\theta_1)+(\hat{\theta}_{0}-\theta_{0})(\hat{\theta}_1-\theta_1)\Big].
\end{align*}
Since $\mathrm{E}[\hat{\theta}_1-\theta_1]=0$, $\mathrm{E}[\hat{\theta}_0-\theta_0]=0$, $\mathrm{E}[\hat{\alpha}_{Y^*}-\alpha_{Y^*}]=0$ and $\mathrm{E}[\hat{\beta}_{Y^*}-\beta_{Y^*}]=0$ an approximation of the expected value of the estimator $\hat{\alpha}_Y$ is given by:
\begin{flalign*}
\mathrm{E}\big[\frac{\hat{\alpha}_{Y^*}-\hat{\theta}_0}{\hat{\theta}_1}\big]&\approx
\frac{\alpha_{Y^*}-\theta_0}{\theta_1}+\frac{1}{\theta_1^2}\Big[\frac{\alpha_{Y^*}-\theta_0}{\theta_1}\mathrm{E}\big[(\hat{\theta}_1-\theta_1)^2\big]&\\
&-\mathrm{E}\big[(\hat{\alpha}_{Y^*}-\alpha_{Y^*})(\hat{\theta}_1-\theta_1)\big]+\mathrm{E}\big[(\hat{\theta}_0-\theta_0)(\hat{\theta}_1-\theta_1)\big]\Big]=&\\
&=\frac{\alpha_{Y^*}-\theta_0}{\theta_1}+\frac{1}{\theta_1^2}\Big[\frac{\alpha_{Y^*}-\theta_0}{\theta_1}\mathrm{Var}\big(\hat{\theta}_1\big)-\mathrm{Cov}\big(\hat{\alpha}_{Y^*},\hat{\theta}_1\big)+\mathrm{Cov}\big(\hat{\theta}_0,\hat{\theta}_1\big)\Big]=&\\
&=\alpha_Y+\frac{1}{\theta_1^2}\Big[\frac{\tau^2[\alpha_Y-\bar{Y}^{(c)}]}{\sum(Y_j^{(c)}-\bar{Y}^{(c)})^2}\Big]. 
\end{flalign*}
Congruently, an approximation of the expected value of the estimator $\hat{\beta}_Y$ is given by:
\begin{flalign*}
\mathrm{E}\big[\frac{\hat{\beta}_{Y^*}}{\hat{\theta}_1}\big]&\approx\frac{\beta_{Y^*}}{\theta_1}+\frac{1}{\theta_1^2}\Big[ \frac{\beta_{Y^*}}{\theta_1}\mathrm{E}\big[(\hat{\theta}_1-\theta_1)^2\big]-\mathrm{E}\big[(\hat{\beta}_{Y^*}-\beta_{Y^*})(\hat{\theta}_1-\theta_1)\big]\Big]=&\\
&=\frac{\beta_{Y^*}}{\theta_1}+\frac{1}{\theta_1^2}\Big[\frac{\beta_{Y^*}}{\theta_1}\mathrm{Var}\big(\hat{\theta}_1\big)\Big]=& \\
&=\beta_Y+\frac{1}{\theta_1^2}\Big[\frac{\tau^2\beta_Y}{\sum(Y_j^{(c)}-\bar{Y}^{(c)})^2}\Big]. 
\end{flalign*}
\\
Only using the first order Taylor expansion of the estimators, approximations of the variance of $\hat{\alpha}_Y$ and $\hat{\beta}_Y$ are respectively:
\bgroup\allowdisplaybreaks
\begin{align*}
\mathrm{Var}\big(\frac{\hat{\alpha}_{Y^*}-\hat{\theta}_0}{\hat{\theta}_1}\big)&\approx \frac{1}{\theta_1^2}\Big[\alpha_Y^2\mathrm{Var}\big(\hat{\theta}_1\big)+\mathrm{Var}\big(\hat{\alpha}_{Y^*}-\hat{\theta}_0\big)-2\alpha_Y\mathrm{Cov}\big(\hat{\theta}_1,\hat{\alpha}_{Y^*}-\hat{\theta}_0\big)\Big]=\\
&=\frac{1}{\theta_1^2}\Big[\alpha_Y^2\mathrm{Var}\big(\hat{\theta}_1\big)+\mathrm{Var}\big(\hat{\alpha}_{Y^*}\big)+\mathrm{Var}\big(\hat{\theta}_0\big)-2\mathrm{Cov}\big(\hat{\alpha}_{Y^*},\hat{\theta}_0\big)\\
&-2\alpha_Y\mathrm{Cov}\big(\hat{\theta}_1,\hat{\alpha}_{Y^*}\big)
+2\alpha_Y\mathrm{Cov}\big(\hat{\theta}_1,\hat{\theta}_0\big)\Big]=\\
&=\frac{1}{\theta_1^2}\Big[\frac{(\theta_1^2\sigma^2+\tau^2)\sum{X_i^2}}{N\sum{(X_i-\bar{X})^2}}+\alpha_Y^2\frac{\tau^2}{\sum(Y_j^{(c)}-\bar{Y}^{(c)})^2}+\frac{\tau^2\sum (Y_j^{(c)})^2}{K\sum(Y_j^{(c)}-\bar{Y}^{(c)})^2}\\
&+2\alpha_Y\frac{-\tau^2\bar{Y}^{(c)}}{\sum(Y_j^{(c)}-\bar{Y}^{(c)})^2}\Big]=\\
&=\frac{1}{\theta_1^2}\Big[\frac{(\theta_1^2\sigma^2+\tau^2)\sum{X_i^2}}{N\sum{(X_i-\bar{x})^2}}+\alpha_Y^2\frac{\tau^2}{\sum(y_j^{(c)}-\bar{y}^{(c)})^2}+\frac{\tau^2(\sum(y_j^{(c)}-\bar{y}^{(c)})^2+K(\bar{y}^{(c)})^2)}{K\sum(y_j^{(c)}-\bar{y}^{(c)})^2}\\
&-2\alpha_Y\frac{\tau^2\bar{y}^{(c)}}{\sum(y_j^{(c)}-\bar{y}^{(c)})^2}\Big]=\\
&=\frac{1}{\theta_1^2}\Big[\frac{(\theta_1^2\sigma^2+\tau^2)\sum{x_i^2}}{N \sum{(x_i-\bar{x})^2}}+\tau^2 \Big( \frac{1}{K}+\frac{(\bar{y}^{(c)}-\alpha_Y)^2}{\sum(y_j^{(c)}-\bar{y}^{(c)})^2}\Big)\Big],\\
\mathrm{Var}\big(\frac{\hat{\beta}_{Y^*}}{\hat{\theta}_1}\big)&\approx\frac{1}{\theta_1^2}\Big[\frac{\theta_1^2\sigma^2+\tau^2}{\sum{(x_i-\bar{x})^2}}+\frac{\beta_Y^2\tau^2}{\sum(y_j^{(c)}-\bar{y}^{(c)})^2}\Big].
\end{align*}
\egroup
\subsubsection{Fieller method}
Assume that $\hat{\beta}_{Y^*}$ and $\hat{\theta}_1$ are normally distributed (note that this assumption is satisfied with large study samples ($N$) and large calibration samples ($K$)). The sum of two normally distributed variables is normally distributed, hence, $\hat{\beta}_{Y^*}-\beta_Y\hat{\theta}_1$ is normally distributed.\\
Furthermore, we have
\begin{equation*}
\text{Var}(\hat{\beta}_{Y^*}-\beta_Y\hat{\theta}_1)=\text{Var}(\hat{\beta}_{Y^*})+\beta_Y^2\text{Var}(\hat{\theta}_1).
\end{equation*}
Where,
\begin{eqnarray*}
\text{Var}(\hat{\beta}_{Y^*})&=&\frac{\theta_1^2\sigma^2+\tau^2}{\sum(x_i-\bar{x})^2}\\
\text{Var}(\hat{\theta}_1)&=&\frac{\tau^2}{\sum(y^{(c)}_j-\bar{y}^{(c)})^2}
\end{eqnarray*}
If we now divide the term $\hat{\beta}_{Y^*}-\beta_Y\hat{\theta}_1$ by its standard deviation, we get:
\begin{equation}\label{eq:app_T0}
T_0=\frac{\hat{\beta}_{Y^*}-\beta_Y\hat{\theta}_1}{\sqrt{\frac{\theta_1^2\sigma^2+\tau^2}{\sum(x_i-\bar{x})^2} + \frac{\tau^2}{\sum(y^{(c)}_j-\bar{y}^{(c)})^2} \beta_Y^2 }}
\end{equation}
We are interested to find the set of $\beta_Y$ values for which the corresponding $T_0$ values lie within the $(1-\alpha)$ quantiles of the $t$-distribution with $N-2$ degrees of freedom (this only holds approximately, see for details \cite{Franz2007Ratios:Use}). Let us denote these values by $t_q$, from (\ref{eq:app_T0}) we have,
\begin{equation*}
(\frac{\tau^2}{\sum(y^{(c)}_j-\bar{y}^{(c)})^2}t_q^2-\hat{\theta}_1^2)\beta_Y^2+2\hat{\beta}_{Y^*}\hat{\theta}_1\beta_Y+(
\frac{\theta_1^2\sigma^2+\tau^2}{\sum(x_i-\bar{x})^2}t_q^2-\hat{\beta}_{Y^*}^2)=0.
\end{equation*}
In the case that $\hat{\theta}_1$ is significantly different from zero at a significance level of $\alpha$ (that is,\\ $\hat{\theta}_1/\sqrt{\frac{\tau^2}{\sum(y^{(c)}_j-\bar{y}^{(c)})^2}}>t_q$), solving this for $\beta_Y$ results in the following $(1-\alpha)$ confidence intervals:
\begin{equation*}
\beta_Y=\frac{-\hat{\beta}_{Y^*}\hat{\theta}_1 \pm \sqrt{\hat{\beta}_{Y^*}^2\hat{\theta}_1^2-(\frac{\tau^2}{\sum(y^{(c)}_j-\bar{y}^{(c)})^2}t_q^2-\hat{\theta}_1^2)(\frac{\theta_1^2\sigma^2+\tau^2}{\sum(x_i-\bar{x})^2}t_q^2-\hat{\beta}_{Y^*}^2)}}{\frac{\tau^2}{\sum(y^{(c)}_j-\bar{y}^{(c)})^2}t_q^2-\hat{\theta}_1^2}.
\end{equation*}
In the other case, the confidence intervals are unbounded, see for more details \cite{Franz2007Ratios:Use}.
\subsection{Differential measurement error}
Obvious estimators for $\alpha_Y$ and $\beta_Y$ are:
\begin{align*}
\hat{\alpha}_Y=(\hat{\alpha}_{Y^*}-\hat{\theta}_{00})/\hat{\theta}_{10} &\textrm{\quad and \quad}
\hat{\beta}_Y=(\hat{\beta}_{Y^*}+\hat{\alpha}_{Y^*}-\hat{\theta}_{01})/\hat{\theta}_{11}-\hat{\alpha}_{Y}.
\end{align*}
These estimators can be approximated with a second order Taylor expansion by:
\begin{align*}
\frac{\hat{\alpha}_{Y^*}-\hat{\theta}_{00}}{\hat{\theta}_{10}}&\approx\frac{\alpha_{Y^*}-\theta_{00}}{\theta_{10}}+\frac{1}{\theta_{10}}\Big[-\frac{\alpha_{Y^*}-\theta_{00}}{\theta_{10}}(\hat{\theta}_{10}-\theta_{10})+(\hat{\alpha}_{Y^*}-\alpha_{Y^*})-(\hat{\theta}_{00}-\theta_{00})\Big]\\
&+\frac{1}{\theta_{11}^2}\Big[ \frac{\alpha_{Y^*}-\theta_{00}}{\theta_{10}}(\hat{\theta}_{10}-\theta_{10})^2-(\hat{\alpha}_{Y^*}-\alpha_{Y^*})(\hat{\theta}_{10}-\theta_{10})+(\hat{\theta}_{00}-\theta_{00})(\hat{\theta}_{10}-\theta_{10})\Big],\\
\frac{\hat{\beta}_{Y^*}-\hat{\theta}_{01}}{\hat{\theta}_{11}}&\approx\frac{\beta_{Y^*}-\theta_{01}}{\theta_{11}}+\frac{1}{\theta_{11}}\Big[-\frac{\beta_{Y^*}-\theta_{01}}{\theta_{11}}(\hat{\theta}_{11}-\theta_{11})+(\hat{\beta}_{Y^*}-\beta_{Y^*})-(\hat{\theta}_{01}-\theta_{01})\Big]\\
&+\frac{1}{\theta_{11}^2}\Big[ \frac{\beta_{Y^*}-\theta_{01}}{\theta_{11}}(\hat{\theta}_{11}-\theta_{11})^2-(\hat{\beta}_{Y^*}-\beta_{Y^*})(\hat{\theta}_{11}-\theta_{11})+(\hat{\theta}_{01}-\theta_{01})(\hat{\theta}_{11}-\theta_{11})\Big],\\
\frac{\hat{\alpha}_{Y^*}}{\hat{\theta}_{11}}&\approx\frac{\alpha_{Y^*}}{\theta_{11}}+\frac{1}{\theta_{11}}\Big[-\frac{\alpha_{Y^*}}{\theta_{11}}(\hat{\theta}_{11}-\theta_{11})+(\hat{\alpha}_{Y^*}-\alpha_{Y^*}))\Big]\\
&+\frac{1}{\theta_{11}^2}\Big[ \frac{\alpha_{Y^*}}{\theta_{11}}(\hat{\theta}_{11}-\theta_{11})^2-(\hat{\alpha}_{Y^*}-\alpha_{Y^*})(\hat{\theta}_{11}-\theta_{11})\Big].
\end{align*}
Congruent to the results for the estimators under systematic measurement error, we can conclude:
\begin{flalign*}
\mathrm{E}\big[\frac{\hat{\alpha}_{Y^*}-\hat{\theta}_{00}}{\hat{\theta}_{10}}\big]&\approx
\alpha_Y+\frac{1}{\theta_{10}^2}\Big[\alpha_Y\text{Var}\big(\hat{\theta}_{10}\big) + \text{Cov}\big(\hat{\theta}_{00},\hat{\theta}_{10}\big)\Big]. 
\end{flalign*}
Congruently, an approximation of the expected value of the estimator $\hat{\beta}_Y$ is given by:
\begin{flalign*}
\mathrm{E}\big[\frac{\hat{\beta}_{Y^*}+\hat{\alpha}_{Y^*}-\hat{\theta}_{01}}{\hat{\theta}_{11}}-\hat{\alpha}_Y\big]&\approx\beta_Y+\frac{1}{\theta_{11}^2}\Big[(\beta_Y + \alpha_Y)\text{Var}\big(\hat{\theta}_{11}\big)+\text{Cov}\big(\hat{\theta}_{01}, \hat{\theta}_{11}\big)\Big]\\&-\frac{1}{\theta_{10}^2}\Big[\alpha_Y\text{Var}\big(\hat{\theta}_{10}\big)+\text{Cov}\big(\hat{\theta}_{00}, \hat{\theta}_{10}\big)\Big]. 
\end{flalign*}
\\
And the variance of the estimators is approximated by:
\begin{align*}
\mathrm{Var}\big(\frac{\hat{\alpha}_{Y^*}-\hat{\theta}_{00}}{\hat{\theta}_{10}}\big)&\approx\
\frac{1}{\theta_{10}^2}\Big[\text{Var}\big(\hat{\alpha}_{Y^*}\big)+\alpha_Y^2\text{Var}\big(\hat{\theta}_{10}\big)+\text{Var}\big(\hat{\theta}_{0
0}\big)+2\alpha_Y\text{Cov}\big(\hat{\theta}_{00},\hat{\theta}_{10}\big)\Big],\\
\mathrm{Var}\big(\frac{\hat{\beta}_{Y^*}+\hat{\alpha}_{Y^*}-\hat{\theta}_{01}}{\hat{\theta}_{11}}-\hat{\alpha}_Y\big)&\approx\frac{1}{\theta_{11}^2}\Big[\big(\beta_Y+\alpha_Y\big)^2\text{Var}\big(\hat{\theta}_{11}\big)+\text{Var}\big(\hat{\beta}_{Y^*}\big)+\text{Var}\big(\hat{\alpha}_{Y^*}\big)\\&+2\text{Cov}\big(\hat{\alpha}_{Y^*},\hat{\beta}_{Y^*}\big)+\text{Var}\big(\hat{\theta}_{01}\big)+2\big(\beta_Y+\alpha_Y\big)\text{Cov}\big(\hat{\theta}_{11},\hat{\theta}_{01}\big)\Big]\\&+\mathrm{Var}\big(\hat{\alpha}_Y\big).
\end{align*}
Note that in the case of differential measurement error, we assume that $\text{Cov}\big(\hat{\theta}_{11},\hat{\theta}_{00})=0$,\\ $\text{Cov}\big(\hat{\theta}_{11},\hat{\theta}_{10})=0$, $\text{Cov}\big(\hat{\theta}_{01},\hat{\theta}_{00})=0$ and $\text{Cov}\big(\hat{\theta}_{01},\hat{\theta}_{10})=0$.
\end{document}